\documentclass[10pt,letterpaper]{article}
\usepackage[margin=2cm]{geometry}

\usepackage{amsmath,amssymb}

\usepackage{changepage}

\usepackage[utf8x]{inputenc}

\usepackage{textcomp,marvosym}

\usepackage{cite}

\usepackage{nameref,hyperref}

\usepackage{microtype}
\DisableLigatures[f]{encoding = *, family = * }

\usepackage[table]{xcolor}

\usepackage{array}

\newcolumntype{+}{!{\vrule width 2pt}}

\newlength\savedwidth

\usepackage[aboveskip=1pt,labelfont=bf,labelsep=period,justification=raggedright,singlelinecheck=off]{caption}

\makeatletter
\renewcommand{\@biblabel}[1]{\quad#1.}
\makeatother

\usepackage{lastpage,fancyhdr,graphicx}
\usepackage{epstopdf}
\fancyheadoffset[L]{2.25in}
\fancyfootoffset[L]{2.25in}
\newcommand{\beginsupplement}{%
        \setcounter{table}{0}
        \renewcommand{\thetable}{S\arabic{table}}%
        \setcounter{figure}{0}
        \renewcommand{\thefigure}{S\arabic{figure}}%
        \setcounter{equation}{0}
        \renewcommand{\theequation}{S\arabic{equation}}
     }

\usepackage{epstopdf}

\usepackage{subfig}

\begin{document}
\vspace*{0.2in}

\begin{flushleft}
{\Large
\textbf\newline{Resist or perish: fate of a microbial population subjected to a periodic presence of antimicrobial} 
}
\newline
\\
Loïc Marrec\textsuperscript{1},
Anne-Florence Bitbol\textsuperscript{1,2*}
\\
\bigskip
\textbf{1} Sorbonne Universit{\'e}, CNRS, Institut de Biologie Paris-Seine, Laboratoire Jean Perrin (UMR 8237), F-75005 Paris, France\\
\textbf{2} Institute of Bioengineering, School of Life Sciences, École Polytechnique Fédérale de Lausanne (EPFL), CH-1015 Lausanne, Switzerland
\bigskip

%
%


* anne-florence.bitbol@epfl.ch

\end{flushleft}
\section*{Abstract}
The evolution of antimicrobial resistance can be strongly affected by variations of antimicrobial concentration. Here, we study the impact of periodic alternations of absence and presence of antimicrobial on resistance evolution in a microbial population, using a stochastic model that includes variations of both population composition and size, and fully incorporates stochastic population extinctions. We show that fast alternations of presence and absence of antimicrobial are inefficient to eradicate the microbial population and strongly favor the establishment of resistance, unless the antimicrobial increases enough the death rate. We further demonstrate that if the period of alternations is longer than a threshold value, the microbial population goes extinct upon the first addition of antimicrobial, if it is not rescued by resistance. We express the probability that the population is eradicated upon the first addition of antimicrobial, assuming rare mutations. Rescue by resistance can happen either if resistant mutants preexist, or if they appear after antimicrobial is added to the environment. Importantly, the latter case is fully prevented by perfect biostatic antimicrobials that completely stop division of sensitive microorganisms. By contrast, we show that the parameter regime where treatment is efficient is larger for biocidal drugs than for biostatic drugs. This sheds light on the respective merits of different antimicrobial modes of action.

\section*{Author summary}
Antimicrobials select for resistance, which threatens to make antimicrobials useless. Understanding the evolution of antimicrobial resistance is therefore of crucial importance. Under what circumstances are microbial populations eradicated by antimicrobials? Conversely, when are they rescued by resistance? We address these questions employing a stochastic model that incorporates variations of both population composition and size. We consider periodic alternations of absence and presence of antimicrobial, which may model a treatment. We find a threshold period above which the first phase with antimicrobial fully determines the fate of the population. Faster alternations strongly select for resistance, and are inefficient to eradicate the microbial population, unless the death rate induced by the treatment is large enough. For longer alternation periods, we calculate the probability that the microbial population gets eradicated. We further demonstrate the different merits of biostatic antimicrobials, which prevent sensitive microbes from dividing, and of biocidal ones, which kill sensitive microbes.


\section*{Introduction}

Antibiotics and antivirals allow many major infectious diseases to be treated. However, with the increasing use of antimicrobials, pathogenic microorganisms tend to become resistant to these drugs, which then become useless. Understanding the evolution of resistance is of paramount importance in order to fight the major public health issue raised by antimicrobial resistance~\cite{WHO,AMR}.

The evolution of antimicrobial resistance often occurs in a variable environment, as antimicrobial is added and removed from a medium or given periodically to a patient~\cite{Lin16,Levin-Reisman17}. This results into varying patterns of selection, which are known to have a dramatic effect on evolution in other contexts~\cite{Mustonen08,Rivoire11,Melbinger15,Desponds16,Wienand17}. To address how variations of antimicrobial concentration impact resistance evolution, we investigate theoretically the \textit{de novo} acquisition of resistance in a microbial population in the presence of alternations of phases of presence and absence of antimicrobial. This situation can represent, for example, a treatment where the concentration within the patient falls under the Minimum Inhibitory Concentration (MIC) between drug intakes~\cite{Regoes04}, which is a realistic case~\cite{Jacobs01,Regoes04}. 

We propose a general stochastic model that incorporates variations of both population composition and size, i.e. population genetics and population dynamics. Despite having a common origin in stochastic birth, death and mutation events, and thus being intrinsically coupled, these phenomena are seldom considered together in theoretical studies~\cite{Melbinger10}. However, it is particularly crucial to address both of them when studying the evolution of antimicrobial resistance, because the aim of an antimicrobial treatment is to eradicate a microbial population, or at least to substantially decrease its size, while the evolution of resistance corresponds to a change in the genetic makeup of the population. Our general model allows us to fully incorporate the stochasticity of mutation occurrence and establishment~\cite{Ewens79,Rouzine01,Fisher07,Patwa08,Weissman09}, as well as that of population extinction, whose practical importance was recently highlighted~\cite{Coates18,Teimouri19,AlexanderPreprint}.

In this framework, we ask whether a microbial population subject to alternations of phases of presence and absence of antimicrobial develops resistance, which corresponds to treatment failure and to rescue of the microbial population by resistance~\cite{Martin12,Alexander14}, or goes extinct, which corresponds to treatment success. In other words, we ask whether the microbial population resists or perishes.

We study both the impact of biocidal drugs, that kill microorganisms, and of biostatic drugs, that prevent microorganisms from growing. We show that fast alternations of phases with and without antimicrobial do not permit eradication of the microbial population before resistant mutants fix, unless the death rate with antimicrobial is large enough. Conversely, intermediate alternation speeds are effective for a wider range of antimicrobial modes of action, but the probability of population extinction and therefore of treatment success, which we fully quantify, is not one, because resistance can rescue the population, and this effect depends on the size of the microbial population. We find that the parameter range where antimicrobial treatment is efficient is larger for biocidal drugs than for biostatic drugs. However, we also show that biocidal and imperfect biostatic antimicrobials permit an additional mechanism of rescue by resistance compared to biostatic drugs that completely stop growth. This sheds light on the respective merits of different antimicrobial modes of action. Finally, we find a population size-dependent critical drug concentration below which antimicrobials cannot eradicate microbial populations.

\section*{Model and methods}  

We consider a microbial population with carrying capacity $K$, corresponding to the maximum population size that the environment can sustain, given the nutrients available. The division rate of each microorganism is assumed to be logistic, and reads $f(1-N/K)$, where $N$ represents the total population size, while the fitness $f$ is the maximal division rate of the microorganism, reached when $N\ll K$. This model therefore incorporates population size variations, and allows us to include extinctions induced by the antimicrobial drug.

Mutations that confer antimicrobial resistance are often associated with a fitness cost, i.e. a slower reproduction~\cite{Borman96,Andersson10,zurWiesch11}, but this fitness cost can be compensated by subsequent mutations~\cite{Schrag97,Levin00,Paulander07,deSousa15}. The acquisition of resistance is therefore often irreversible, even if the antimicrobial is removed from the environment~\cite{Schrag97,Andersson10}. Thus motivated, we consider three types of microorganisms: sensitive (S) microorganisms, whose division or death rate is affected by antimicrobials, resistant (R) microorganisms, that are not affected by antimicrobials but that bear a fitness cost, and resistant-compensated (C) microorganisms that are not affected by antimicrobials and do not bear a fitness cost. In the absence of antimicrobial, their fitnesses (maximal division rates) are denoted by $f_S$, $f_R$ and $f_C$, respectively, and their death rates by $g_S$, $g_R$ and $g_C$. Values in the presence of antimicrobial are denoted by a prime, e.g. $f'_S$. Note that we include small but nonzero baseline death rates, which can model losses or the impact of the immune system \textit{in vivo}, and allows for population evolution even at steady-state size. Without loss of generality, we set $f_S=1$ throughout. In other words, the maximum reproduction rate of S microorganisms, attained when population size is much smaller than the carrying capacity, sets our time unit. We further denote by $\mu_1$ and $\mu_2$ the mutation probabilities upon each division for the mutation from S to R and from R to C, respectively. In several actual cases, the effective mutation rate towards compensation is higher than the one towards the return to sensitivity, because multiple mutations can compensate for the initial cost of resistance~\cite{Levin00,Paulander07,Hughes15}. Thus, we do not take into account back-mutations. Still because of the abundance of possible compensatory mutations, often  $\mu_1\ll\mu_2$~\cite{Levin00,Poon05}. We provide general analytical results as a function of $\mu_1$ and $\mu_2$, and we focus more on the limit $\mu_1\ll\mu_2$, especially in simulations. 

Our model thus incorporates both population dynamics and population genetics~\cite{Melbinger10,Melbinger15,Huang15}, and is more realistic than descriptions assuming constant population sizes~\cite{Marrec18}, e.g. in the framework of the Moran process~\cite{Moran58,Ewens79}. Throughout, our time unit corresponds to a generation of sensitive microorganisms without antimicrobial in the exponential phase (reached when $N\ll K$). 

The action of an antimicrobial drug can be quantified by its MIC, which corresponds the minimum concentration that stops the growth of a microbial population~\cite{Andersson10}. More precisely, the MIC corresponds to the concentration such that death rate and division rate are equal~\cite{Coates18}: in a deterministic framework, above the MIC, the population goes extinct, while below it, it grows until reaching carrying capacity. We investigate the impact of periodic alternations of phases of absence and presence of antimicrobial, at concentrations both above and below the MIC. We consider both biostatic antimicrobials, which decrease the division rate of microorganisms ($f'_S<f_S$), and biocidal antimicrobials, which increase the death rate of microorganisms ($g'_S>g_S$)~\cite{Coates18}.

We start from a microbial population where all individuals are S (sensitive), without antimicrobial. Specifically, we generally start our simulations with 10 S microorganisms, thus including a phase of initial growth, which can model the development of an infection starting from the bottleneck at transmission~\cite{Abel15}. Our results are robust to variations of this initial condition, since we mainly consider timescales longer than that of the initial growth of the population to its equilibrium size. Note however that if we started with a very small number of S microorganisms (i.e. 1 or 2), we would need to take into account rapid stochastic extinctions (see Fig. \ref{quick_ext_plot}B). 

Antimicrobial both drives the decrease of the population of sensitive microorganisms and selects for resistance. We ask whether resistance fully evolves \textit{de novo}, leading to the C microorganisms taking over, or whether the microbial population goes extinct before this happens. The first case corresponds to treatment failure, and the second to treatment success. Hence, we are interested in the probability $p_0$ of extinction of the microbial population before C microorganisms fix in the population, i.e. take over. We also discuss the average time $t_{fix}$ it takes for the population to fully evolve resistance, up to full fixation of the C microorganisms, and the mean time to extinction before the fixation of the C type $t_{ext}$.  

We present both analytical and numerical results. Our analytical results are obtained using methods from stochastic processes, including the Moran process at fixed population size~\cite{Ewens79} and birth-death processes with time varying rates~\cite{NissenMeyer66,Bailey,Alexander12,Parzen}. Our simulations employ a Gillespie algorithm~\cite{Gillespie76,Gillespie77}, and incorporate all individual stochastic division, mutation and death events with their exact rates (see Supporting Information, section~\ref{SI_Simu} for details).

\section*{Results}

\subsection*{Conditions for a periodic presence of perfect biostatic antimicrobial to eradicate the microbial population}

\begin{figure}[htb]%
	\centering
	\includegraphics[width=1.1\textwidth]{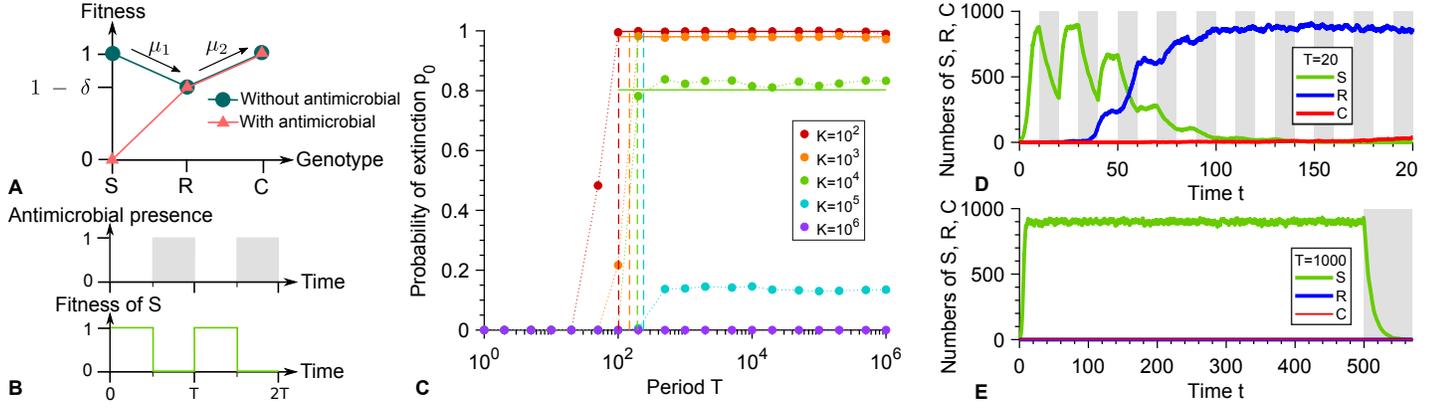}
	\vspace{0.2cm}
	\caption{{\bf Periodic presence of a perfect biostatic antimicrobial.} \textbf{A:} Microbial fitness versus genotype with and without antimicrobial. Genotypes are the following: S: sensitive; R: resistant; C: resistant-compensated. $\delta$ represents the fitness cost of resistance. \textbf{B:} Periodic presence of antimicrobial (gray: presence, white: absence), and impact on the fitness of S microorganisms. \textbf{C:} Probability $p_0$ that the microbial population goes extinct before resistance gets established versus alternation period $T$, for various carrying capacities $K$. Markers: simulation results, with probabilities estimated over $10^2 - 10^3$ realizations. Horizontal solid lines: analytical predictions from Eq.~\ref{p0_biostatic}. Dashed lines: $T/2=\tau_S$. \textbf{D and E:} Numbers of S, R and C microorganisms versus time in example simulation runs for $K=1000$, with $T=20$ and $T=1000$ respectively. In \textbf{D}, resistance takes over, while in \textbf{E}, extinction occurs shortly after antimicrobial is first added. Phases without (resp. with) antimicrobial are shaded in white (resp. gray). Parameter values: $f_S=1$ without antimicrobial, $f'_S=0$ with antimicrobial, $f_R=0.9$, $f_C=1$, $g_S=g_R=g_C=0.1$, $\mu_1=10^{-5}$ and $\mu_2=10^{-3}$. All simulations start with 10 S microorganisms. }%
	\label{Per_pres_bios}%
\end{figure}

Do periodic alternations of phases with and without antimicrobial allow the eradication of a microbial population, or does resistance develop? We first address this question in the case of a biostatic antimicrobial sufficiently above the MIC to completely stop the growth of S microorganisms (see Fig.~\ref{Per_pres_bios}A-B). With such a ``perfect'' biostatic antimicrobial, the fitness of S microorganisms is $f'_S=0$, while without antimicrobial, $f_S=1$. Here, we assume that the death rate of S microorganisms is not affected by the antimicrobial, i.e. $g'_S=g_S$, but the case of a biocidal antimicrobial will be considered next. Note that within our logistic growth model, we consider that S microorganisms that cannot divide still consume resources, e.g. nutrients, in order to self-maintain. They may also still grow in size even if they cannot divide~\cite{Lin16}.

A crucial point is how the duration of a phase with antimicrobial, which corresponds here to the half-period $T/2$ of alternations, compares to the average time $\tau_S$ needed for a population of S microorganisms to go extinct in the presence of antimicrobial. Indeed, if $T/2\gg \tau_S$, one single phase with antimicrobial suffices to eradicate a microbial population in the absence of resistance. An exact first passage time calculation~\cite{Marrec18,Sekimoto10} (see Supporting Information, section~\ref{SI_mfpt}, Eq.~\ref{t_j0_a}) yields $\tau_S=(1/g_S)\times\sum_{i=1}^{N}(1/i) \approx \log(N)/g_S$, where $N\gg1$ represents the number of microorganisms when antimicrobial is first added, i.e. at $T/2$. If the phase before antimicrobial is added is much longer than the initial growth timescale of the population, i.e. if $T/2\gg 1/(f_S-g_S)$ (see Supporting Information, section~\ref{tr_SI}), $N$ can be taken equal to the deterministic equilibrium population size $N=K(1-g_S/f_S)$, obtained by setting the birth rate $f_S(1-N/K)$ equal to the death rate $g_S$. Hence, $\tau_S\approx\log\left[K(1-g_S/f_S)\right]/g_S$. Note that in this regime, the initial population size has no impact on $\tau_S$, and that the division and death rates are both given by $g_S$.  Our simulation results in Fig.~\ref{Per_pres_bios}C display an abrupt increase in the probability $p_0$ that the microbial population goes extinct before developing resistance for $T=2\tau_S$, in good agreement with our analytical prediction.

\begin{figure}[h!]%
	\centering
	\includegraphics[width=\textwidth]{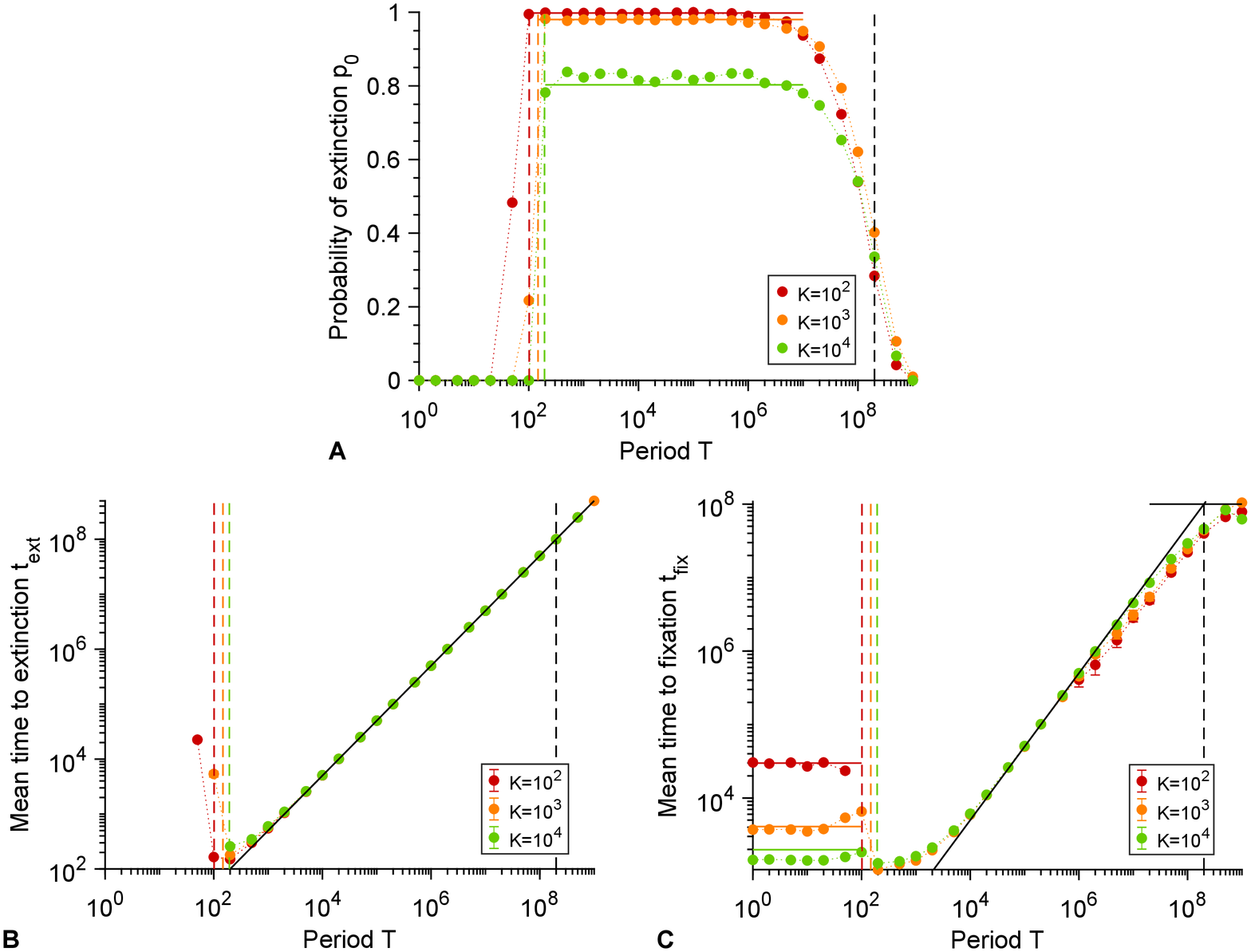}
	\vspace{0.2cm}
	\caption{{\bf Periodic presence of a biostatic antimicrobial that fully stops growth, including long periods.} \textbf{A:} Probability $p_0$ that the microbial population goes extinct before resistance gets established versus alternation period $T$, for various carrying capacities $K$. Markers: simulation results, with probabilities estimated over $10^2 - 10^3$ realizations. Horizontal solid colored lines: analytical predictions from Eq.~\ref{p0_biostatic}. Horizontal solid black line: average spontaneous valley crossing time $\tau_V=(f_S-f_R)/(\mu_1\mu_2g_S)$ (see main text).  \textbf{B:} Average time $t_{ext}$ to extinction versus alternation period $T$ for various carrying capacities $K$. Data shown if extinction occurred in at least 10 realizations. \textbf{C:} Average time  $t_{fix}$ to fixation of the C microorganisms versus alternation period $T$ for various carrying capacities $K$. Data shown if resistance took over in at least 10 realizations. Horizontal solid lines: analytical predictions for very small $T$, using the self-averaged fitness $\tilde{f}_S$ (see main text). In panels \textbf{B} and \textbf{C}, markers are averages over $10^2 - 10^3$ simulation realizations, error bars (often smaller than markers) represent $95\%$ confidence intervals, and the oblique black line corresponds to $T/2$. In all panels, colored dashed lines correspond to $T/2=\tau_S$, while black dashed lines correspond to $T/2=\tau_V$. Parameter values: $f_S=1$ without antimicrobial, $f'_S=0$ with antimicrobial, $f_R=0.9$, $f_C=1$, $g_S=g_R=g_C=0.1$, $\mu_1=10^{-5}$ and $\mu_2=10^{-3}$. All simulations start with 10 S microorganisms. }%
	\label{Per_pres_bios_vc}%
\end{figure}

For fast alternations satisfying $T/2 \ll \tau_S$, the phases with antimicrobial are not long enough to eradicate the microbial population, yielding a \textit{systematic evolution of resistance}, and thus a vanishing probability $p_0$ of extinction before resistance takes over. This prediction is confirmed by our simulation results in Fig.~\ref{Per_pres_bios}C, and an example of resistance evolution in this regime is shown in Fig.~\ref{Per_pres_bios}D. In the limit of very fast alternations, we expect an effective averaging of the fitness of S microorganisms, with $\tilde{f}_S=0.5$. Thus, an R mutant whose lineage will take over the population (i.e. fix) appears after an average time $\tilde{t}_R^a=1/(\tilde{N}\mu_1g_S \tilde{p}_{SR})$ where $\tilde{N}\mu_1g_S$ represents the total mutation rate in the population, with $\tilde{N}=K(1-g_S/\tilde{f}_S)$, and where $\tilde{p}_{SR}=(1-\tilde{f}_S/f_R)/[1-(\tilde{f}_S/f_R)^{\tilde{N}}]$ is the probability that a single R mutant fixes in a population of S microorganisms with constant size $\tilde{N}$, calculated within the Moran model~\cite{Ewens79}. Note that when the effective fitness of S microorganisms is $\tilde{f}_S$, acquiring resistance is beneficial (provided that the fitness cost of resistance is reasonable, namely smaller than 0.5). Subsequently, C mutants will appear and fix, thus leading to the full evolution of resistance in the population. The corresponding average total time $t_{fix}$ of resistance evolution~\cite{Marrec18} obtained in our simulations agrees well with the analytical expression of $\tilde{t}_R^a$ for $T/2 \ll \tau_S$ (see Fig.~\ref{Per_pres_bios_vc}C).

Conversely, if $T/2\gg\tau_S$, \textit{the microbial population is eradicated by the first phase with antimicrobial, provided that no resistant mutant preexists when antimicrobial is added to the environment}. Indeed, resistance cannot appear in the presence of a perfect biostatic antimicrobial since S microorganisms then cannot divide. Thus, in the absence of existing R mutants, extinction occurs shortly after time $T/2$ (see Fig.~\ref{Per_pres_bios_vc}B), and the situation is equivalent to adding antimicrobial at $T/2$ and leaving it thereafter, as exemplified by Fig.~\ref{Per_pres_bios}E. Hence, while they are longer than those usually encountered in periodic treatments, the longest periods considered here are relevant to describe extended continuous treatments. Note that although unlikely, fixation of resistance in the absence of antimicrobial will end up happening by spontaneous fitness valley crossing if the first phase without antimicrobial is long enough. Specifically, this will occur if $T/2 \gg \tau_V$, where $\tau_V\approx(f_S-f_R)/(\mu_1\mu_2g_S)$ is the average valley crossing time by tunneling, which is the relevant process unless populations are very small~\cite{Nowak02,Weinreich05,Weissman09,Marrec18}. Accordingly, our simulation results in Fig. \ref{Per_pres_bios_vc}, which includes longer alternation periods than Fig. \ref{Per_pres_bios}, feature three distinct regimes, and vanishing extinction probabilities are obtained for $T/2 \gg \tau_V$, as well as for $T/2 \ll \tau_S$. 

Let us now focus on the regime where antimicrobial treatment can induce extinction of the microbial population, namely $\tau_S\ll T/2 \ll \tau_V$, and calculate the extinction probability $p_0$. A necessary condition for the population to be rescued by resistance~\cite{Martin12} and avoid extinction is that at least one R mutant be present when antimicrobial is added. In the rare mutation regime $K\mu_1\ll 1$, this occurs with probability $p_R=\tau_R^d/t_R^{app}=N\mu_1 g_S \tau_R^d$, where $t_R^{app}=1/(N\mu_1 g_S)$ is the average time of appearance of a resistant mutant, while $\tau_R^d$ is the average lifetime of a resistant lineage (destined for extinction without antimicrobial), both calculated in a population of S individuals with fixed size $N=K(1-g_S/f_S)$~\cite{Ewens79,Marrec18}. Importantly, the presence of R mutants does not guarantee the rescue of the microbial population, because small subpopulations of R microorganisms may undergo a rapid stochastic extinction. The probability $p_R^e(i)$ of such an extinction event depends on the number of R microorganisms present when antimicrobial is added, which is $i$ with a probability denoted by $p_R^c(i)$, provided that at least one R mutant is present. The probability $p_0$ that the microbial population is not rescued by resistance and goes extinct can then be expressed as: 
\begin{equation}
p_0=1-p_R\sum_{i=1}^{N-1}p_R^c(i)(1-p_R^e(i))\,.
\label{p0_biostatic}
\end{equation}
The probability $p_R^c(i)$ can be calculated within the Moran model since the population size is stable around $N=K(1-g_S/f_S)$ before antimicrobial is added. Specifically, it can be expressed as the ratio of the average time $\tau_{R,i}^d$ the lineage spends in the state where $i$ mutants exist to the total lifetime $\tau_R^d$ of the lineage without antimicrobial: $p_R^c(i)=\tau_{R,i}^d/\tau_R^d$ (see Supporting Information, section~\ref{pRc_SI}). Next, in order to calculate the probability $p_R^e(i)$ that the lineage of R mutants then quickly goes extinct, we approximate the reproduction rate of the R microorganisms by $f_R(1-(S(t)+R(t))/K)\approx f_R(1-S(t)/K)$,  where $S(t)$ and $R(t)$ are the numbers of S and R individuals at time $t$. Indeed, early extinctions of R mutants tend to happen shortly after the addition of antimicrobials, when $S(t)\gg R(t)$. Thus motivated, we further take the deterministic approximation $S(t)=K(1-g_S/f_S)e^{-g_St}$, while retaining a stochastic description for the R mutants~\cite{NissenMeyer66,Bailey}. We then employ the probability generating function $\phi_{i}(z,t)=\sum_{j=0}^{\infty}z^jP(j,t|i,0)$, where $i$ is the initial number of R microorganisms, which satisfies $p_R^e(i)=\lim_{t\to\infty}P(0,t|i,0)=\lim_{t\to\infty}\phi_{i}(0,t)$. Solving the partial differential equation governing the evolution of $\phi_{i}(z,t)$ (see  Supporting Information, section~\ref{pRe_SI}) yields~\cite{Alexander12,Parzen}
\begin{equation}
p_R^e(i)=\lim_{t\to\infty}\left[\frac{g_R\int_0^{t}e^{\rho(u)}du}{1+g_R\int_0^{t}e^{\rho(u)}du}\right]^{i}\,,
\label{pRe_bios}
\end{equation}
with
	\begin{equation}
	\rho(t)=\int_{0}^t\left[g_R-f_R\left(1-\frac{S(u)}{K}\right)\right]du\mbox{ }.
	\end{equation} 
Eq.~\ref{p0_biostatic} then allows us to predict the probability that the microbial population goes extinct thanks to the first addition of antimicrobial. Fig.~\ref{Per_pres_bios}C demonstrates a very good agreement between this analytical prediction and our simulation results in the rare mutation regime $K\mu_1\ll 1$, and Fig.~\ref{pR_pRc_pRe} further demonstrates good agreement for each separate term of Eq.~\ref{p0_biostatic} in this regime. For larger populations, the probability that the microbial population is rescued by resistance increases, and the extinction probability tends to zero for frequent mutations $K\mu_1\gg 1$ because R mutants are then always present in the population, in numbers that essentially ensure their survival (see Fig.~\ref{Per_pres_bios}C). Note that in our simulations presented in Fig.~\ref{Per_pres_bios}, we chose $\mu_1=10^{-5}$ for tractability. With realistic bacterial mutation probabilities, namely $\mu_1\sim 10^{-10}$~\cite{Wielgoss11}, the rare mutation regime remains relevant for much larger populations.

\subsection*{Biocidal antimicrobials and imperfect biostatic ones allow an extra mechanism of rescue by resistance}

\begin{figure}%
	\centering
	\includegraphics[width=\textwidth]{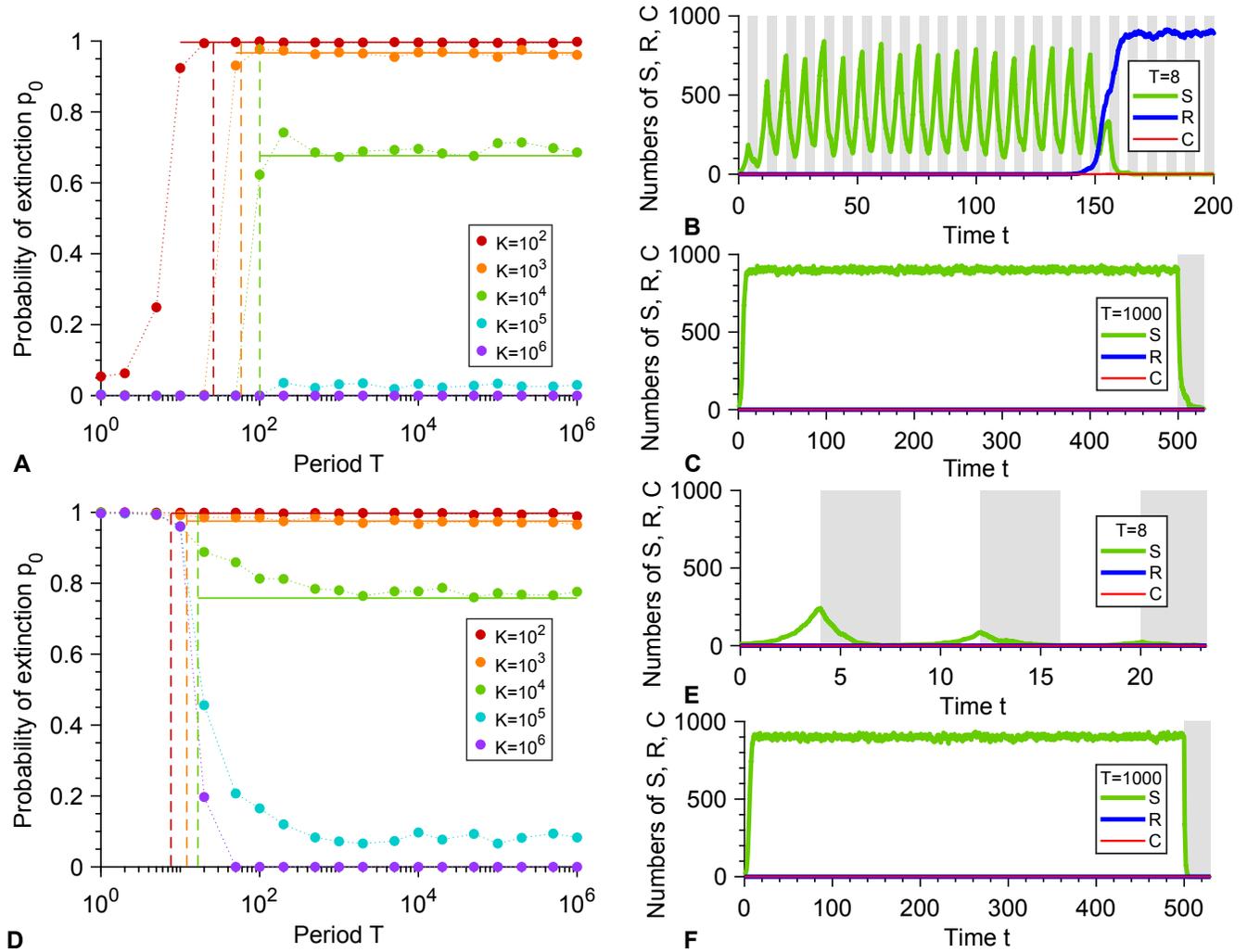}
	\vspace{0.2cm}
\caption{{\bf Periodic presence of a biocidal antimicrobial above the MIC.} \textbf{A:} Probability $p_0$ that the microbial population goes extinct before resistance gets established versus alternation period $T$, for various carrying capacities $K$. Markers: simulation results, with probabilities estimated over $10^2 - 10^3$ realizations. Horizontal solid lines: analytical predictions from Eq.~\ref{p0_biocidal}. Dashed lines: $T/2=\tau_S$. \textbf{B} and \textbf{C:} Numbers of sensitive (S), resistant (R) and compensated (C) microorganisms versus time in example simulation runs for $K=1000$, with $T=8$ and $T=1000$ respectively. In \textbf{B}, resistance takes over, while in \textbf{C}, extinction occurs shortly after antimicrobial is first added. Phases without (resp. with) antimicrobial are shaded in white (resp. gray). Parameter values in \textbf{A}, \textbf{B} and \textbf{C}: $f_S=1$, $f_R=0.9$, $f_C=1$, $g_S=0.1$ without antimicrobial, $g'_S=1.1$ with antimicrobial, $g_R=g_C=0.1$, $\mu_1=10^{-5}$ and $\mu_2=10^{-3}$. All simulations start with 10 S microorganisms. \textbf{D}, \textbf{E} and \textbf{F}: same as \textbf{A}, \textbf{B} and \textbf{C}, but with $g'_S=2$. All other parameters are the same.
	}%
	\label{Per_pres_bioc}%
\end{figure}

How does the mode of action of the antimicrobial impact our results? So far, we considered a perfect biostatic antimicrobial that stops the growth of sensitive microorganisms but does not affect their death rate. Let us now turn to the general case of an antimicrobial that can affect both the division rate and the death rate of sensitive microorganisms, and let us assume that we are above the MIC, i.e. $g'_S>f'_S$. In this section, we present general calculations, but focus most of our discussion on purely biocidal antimicrobials, which increase the death rate of sensitive microorganisms without affecting their growth rate, and compare them to purely biostatic antimicrobials. 
Again, a crucial point is how the duration $T/2$ of a phase with antimicrobial compares to the average time $\tau_S$ needed for a population of S microorganisms to go extinct in the presence of antimicrobial (see Eq.~\ref{tj0}). Indeed, our simulation results in Figs.~\ref{Per_pres_bioc}A and~\ref{Per_pres_bioc}D display an abrupt change in the probability that the microbial population goes extinct before developing resistance for $T=2\tau_S$.

For small periods $T/2\ll \tau_S$, one phase with antimicrobial is not long enough to eradicate the microbial population. However, \textit{the alternations may induce an overall decrease in the population over multiple periods, then leading to extinction}. This is the case when the deterministic growth timescale $1/(f_S-g_S)$ is larger than the decay timescale $1/(g'_S-f'_S)$. Equivalently, in the limit of very fast alternations, there is no nonzero stationary population size when $\tilde{f}_S=(f_S+f'_S)/2<\tilde{g}_S=(g_S+g'_S)/2$, yielding the same condition. For a biostatic drug such that $g'_S=g_S$, this situation cannot happen if $g_S<f_S/2$, which is realistic since baseline death rates are usually small. Conversely, for a biocidal drug such that $f'_S=f_S$, a systematic evolution of resistance will occur if $g'_S<2f_S-g_S$, while population decay over several periods and extinction will occur if $g'_S>2f_S-g_S$. These predictions are confirmed by the simulation results in Figs.~\ref{Per_pres_bioc}A and D, respectively, and the two different cases are exemplified in Figs.~\ref{Per_pres_bioc}B and E. Both of these regimes can arise, depending on the concentration of biocidal antimicrobial. Figs.~\ref{Per_pres_bioc}A-C corresponds to concentrations just above the MIC, while Figs.~\ref{Per_pres_bioc}D-F correspond to larger concentrations of bactericidal drugs, which can induce death rates equal to several times the birth rate~\cite{Levin10,Khan15}. Note that in Fig.~\ref{Per_pres_bioc}A, the extinction probability is not zero for small periods with $K=10^2$: this is because stochastic extinctions can occur before resistance takes over for such a small equilibrium population size.

\begin{figure}%
	\centering
	\includegraphics[width=\textwidth]{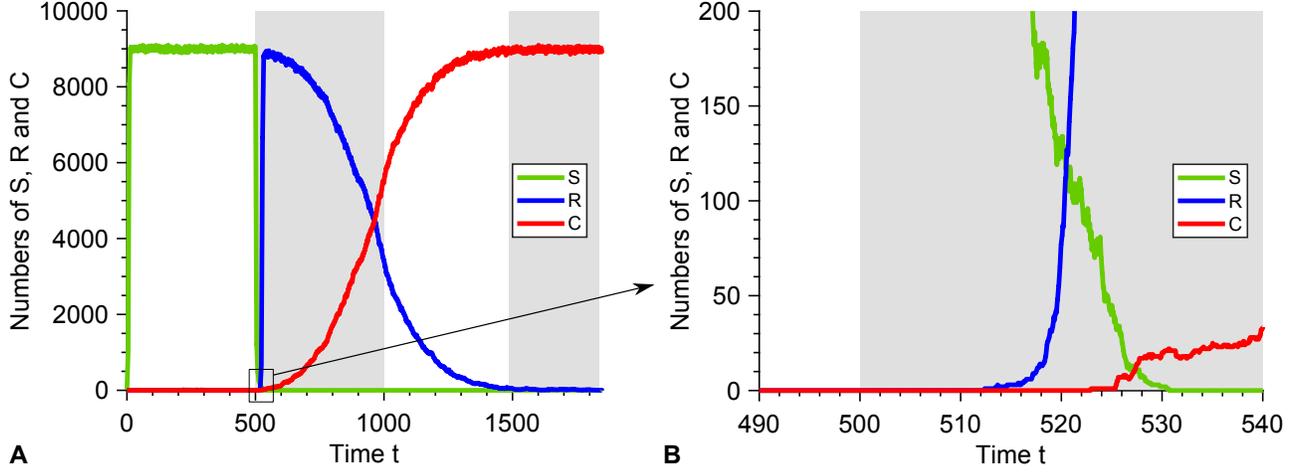}
	\vspace{0.2cm}
\caption{{\bf Resistance emergence in the presence of a biocidal antimicrobial above the MIC.} \textbf{A:} Numbers of sensitive (S), resistant (R) and compensated (C) microorganisms versus time in an example simulation run for $K=10^4$, with $T=1000$. Here resistance takes over. Phases without (resp. with) antimicrobial are shaded in white (resp. gray). \textbf{B:} Zoom showing the emergence of resistance in this realization: an R mutant appears after antimicrobial is added (gray). At this time, the S population is decreasing due to the antimicrobial-induced high death rate, but the surviving S microorganisms are still able to divide. Parameter values  and initial conditions are the same as in Fig.~\ref{Per_pres_bioc}A, B and C.
	}%
	\label{bioc_special}%
\end{figure}

For slower alternations satisfying $T/2 \gg \tau_S$, \textit{the microbial population is eradicated by the first phase with antimicrobial, unless resistance rescues it}. Extinction then occurs shortly after time $T/2$ (see Fig.~\ref{Per_pres_bioc_vc}B and examples in Fig.~\ref{Per_pres_bioc}C and F). Importantly, with a biocidal antimicrobial or with an imperfect biostatic one, \textit{the microbial population can be rescued by resistance in two different ways: either if resistant bacteria are present when antimicrobial is added, or if they appear afterwards}. This second case is exemplified in Fig.~\ref{bioc_special}. It can happen because even at high concentration, such antimicrobials do not prevent S microorganisms from dividing, contrarily to a perfect biostatic one.  Because of this, rescue by resistance can become more likely than with perfect biostatic antimicrobials. Note that, as in the perfect biostatic case, the spontaneous fixation of resistant mutants without antimicrobial will occur if $T/2 \gg \tau_V\approx(f_S-f_R)/(\mu_1\mu_2g_S)$ (see Fig. \ref{Per_pres_bioc_vc}).

Let us focus on the regime where the treatment can efficiently induce extinction, namely $\tau_S\ll T/2 \ll \tau_V$. The probability $p_0$ that the microbial population is not rescued by resistance and goes extinct can then be expressed as:
\begin{equation}
 p_0=\left[1-p_R\sum_{i=1}^{N-1}p_R^c(i)(1-p_R^e(i))\right]\left[1-p_R^a(1-p_R^{e'})\right]\,.
 \label{p0_biocidal}
\end{equation} 
Apart from the last term, which corresponds to resistance appearing after antimicrobial is first added, Eq.~\ref{p0_biocidal} is identical to Eq.~\ref{p0_biostatic}. The quantities $p_R$ and $p_R^c(i)$ are the same as in that case, since they only depend on what happens just before antimicrobial is added. While $p_R^e(i)$ is conceptually similar to the perfect biostatic case, it depends on $f'_S$ and $g'_S$, and its general calculation is presented in Section~\ref{pRe_SI} of the Supporting Information. This leaves us with the new case where resistance appears in the presence of antimicrobial. In the rare mutation regime such that $N_{div}\mu_1 \ll 1$, it happens with probability $p_R^a=N_{div}\mu_1$, where
	\begin{equation}
	N_{div}=\int_0^{\tau_S}f_S\left(1-\frac{S(t)}{K}\right)S(t)\,dt 
	\label{Ndiv}
	\end{equation}
is the number of divisions that would occur in a population of S microorganisms between the addition of antimicrobial (taken as the origin of time here) and extinction. Employing the deterministic approximation for the number $S(t)$ of S microorganisms (see Eq.~\ref{detg}), the probability that the lineage of an R mutant that appears at time $t_0$ quickly goes extinct can be obtained in a similar way as for Eq.~\ref{pRe_bios}, yielding 
	\begin{equation}
p_R^{e'}(t_0)=\lim_{t\to\infty}\frac{g_R\int_{t_0}^{t}e^{\eta(u)}du}{1+g_R\int_{t_0}^{t}e^{\eta(u)}du}\,,
\end{equation}
with
	\begin{equation}
	\eta(t)=\int_{t_0}^t\left[g_R-f_R\left(1-\frac{S(u)}{K}\right)\right]du\mbox{ }.
	\end{equation} 
We then estimate the probability $p_R^{e'}$ that the lineage of an R mutant that appears after the addition of antimicrobial quickly goes extinct by averaging $p_R^{e'}(t_0)$ over the time $t_0$ of appearance of the mutant, under the assumption that exactly one R mutant appears: 
	\begin{equation}
	p_R^{e'}=\int_0^\infty p_R^{e'}(t_0)\,\,\wp_R^{a}(t_0) \,dt_0\,,
	\end{equation}
with	
		\begin{equation}
		\wp_R^{a}(t_0)=\frac{S(t_0)\left(1-\frac{S(t_0)}{K}\right)}{\int_0^\infty S(t)\left(1-\frac{S(t)}{K}\right) dt}\,.
		\label{pdf_pRa}
		\end{equation}
Eq.~\ref{p0_biocidal} then yields the probability that the microbial population goes extinct thanks to the first addition of antimicrobial. Fig.~\ref{Per_pres_bioc}A demonstrates a very good agreement between this analytical prediction and our simulation results in the rare mutation regime $K\mu_1\ll 1$, and Figs.~\ref{pR_pRc_pRe}A-B,~\ref{tauS_Ndiv_pdf} and~\ref{pRe_pRa_pi} further demonstrate good agreement for each term involved in Eq.~\ref{p0_biocidal} in this regime. 

\begin{figure}%
	\centering
	\includegraphics[width=0.6\textwidth]{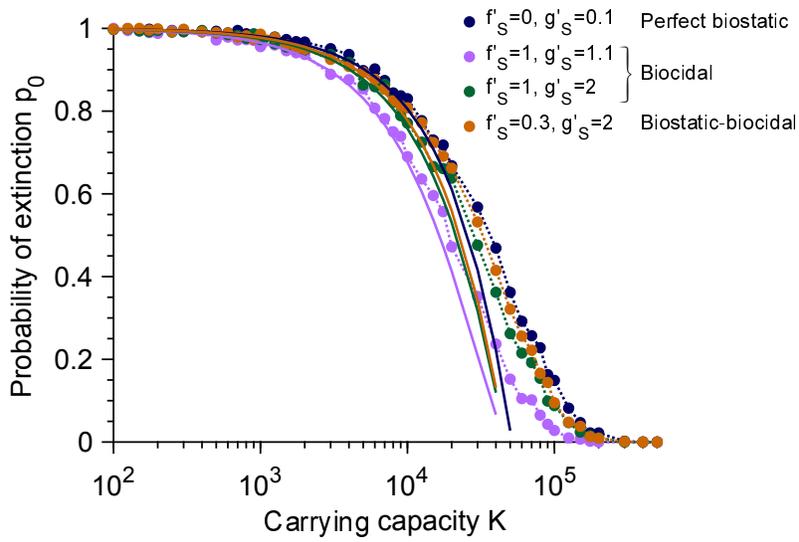}
	\vspace{0.2cm}
\caption{{\bf Dependence of the extinction probability $p_0$ on population size and antimicrobial mode of action.}  The extinction probability $p_0$ is plotted versus carrying capacity $K$ for the perfect biostatic drug (corresponding to Fig.~\ref{Per_pres_bios}), two different concentrations of biocidal drugs yielding two different death rates $g'_S$ (corresponding to Fig.~\ref{Per_pres_bioc}) and a drug with both biostatic and biocidal effects. Markers correspond to simulation results, computed over $10^3$ realizations. Solid lines correspond to our analytical predictions from Eqs.~\ref{p0_biostatic} and~\ref{p0_biocidal}, respectively, which hold for $K\ll 1/\mu_1$. Parameter values  and initial conditions are the same as in Figs.~\ref{Per_pres_bios} and~\ref{Per_pres_bioc}, respectively, and the period of alternations is $T=10^3$, which is in the large-period regime. }%
	\label{p0_vs_K}%
\end{figure}

The extinction probability $p_0$ depends on the size of the microbial population through its carrying capacity $K$ and on the division and death rates with antimicrobial. Fig.~\ref{p0_vs_K} shows the decrease of $p_0$ with $K$, with $p_0$ reaching 0 for $K\mu_1\gg 1$ since resistant mutants are then always present when antimicrobial is added. Moreover, Fig.~\ref{p0_vs_K} shows that $p_0$ \textit{depends on the antimicrobial mode of action}, with large death rates favoring larger $p_0$ in the biocidal case, and with the perfect biostatic antimicrobial yielding the largest $p_0$. Qualitatively, the observed increase of $p_0$ as $g'_S$ increases with a biocidal drug mainly arises from the faster decay of the population of S microorganisms, which reduces the probability $p_R^a$ that an R mutant appears in the presence of antimicrobial. Furthermore, one can show that the extinction probability $p_0$ is larger for a perfect biostatic antimicrobial than for a perfect biocidal antimicrobial with $g'_S\rightarrow\infty$ (see Supporting Information, Section~\ref{Perfectperfect}). Indeed, S microorganisms survive longer in the presence of a perfect biostatic drug, which reduces the division rate of the R mutants due to the logistic growth term, and thus favors their extinction. Such a competition effect is realistic if S microorganisms still take up resources (e.g. nutrients) even while they are not dividing. Besides, a treatment combining biostatic and biocidal effects yields a larger $p_0$ than a pure biocidal one inducing the same death rate, thereby illustrating the interest of the additional biostatic effect (see Fig.~\ref{p0_vs_K}). Note that conversely, adding a biocidal to a perfect biostatic slightly decreases $p_0$ due to the competition effect, as S microorganisms go extinct faster than with the perfect biostatic drug alone.

\clearpage

\subsection*{Sub-MIC drug concentrations and stochastic extinctions}

So far, we considered antimicrobial drugs above the MIC, allowing deterministic extinction in the absence of resistance for long enough drug exposure times. However, sub-MIC drugs can also have a major impact on the evolution of resistance, by selecting for resistance without killing large microbial populations, and moreover by facilitating stochastic extinctions in finite-sized microbial populations~\cite{Coates18,Teimouri19,AlexanderPreprint}. In the sub-MIC regime where $f'_S>g'_S$, the population has a nonzero deterministic equilibrium size $N'=K(1-g'_S/f'_S)$ in the presence of antimicrobial. Nevertheless, stochastic extinctions can remain relatively fast, especially in the weakly-sub-MIC regime where $f'_S$ is close to $g'_S$, and if $K$ is not very large. The key point is whether resistance appears before the extinction time $\tau_S$. The average time of appearance of an R mutant that fixes in a population of $N'$ individuals in the presence of sub-MIC antimicrobial is $t_R^a=1/(N' \mu_1 g'_S p'_{SR})$, where $p'_{SR}=[1-f'_S g_R/(f_R g'_S)]/[1-(f'_S g_R/(f_R g'_S))^{N'}]$ is the fixation probability of an R mutant in a population of S individuals with fixed size $N'$ (see Supporting Information, Section~\ref{Sec_PSR}, and Ref.~\cite{Traulsen09}). Therefore, we expect resistance to take over and the extinction probability $p_0$ to be very small if $t_R^a\ll\tau_S$ below the MIC, even for large periods such that $\tau_S<T/2$.  
 
\begin{figure}[htb]%
	\centering
	\includegraphics[width=\textwidth]{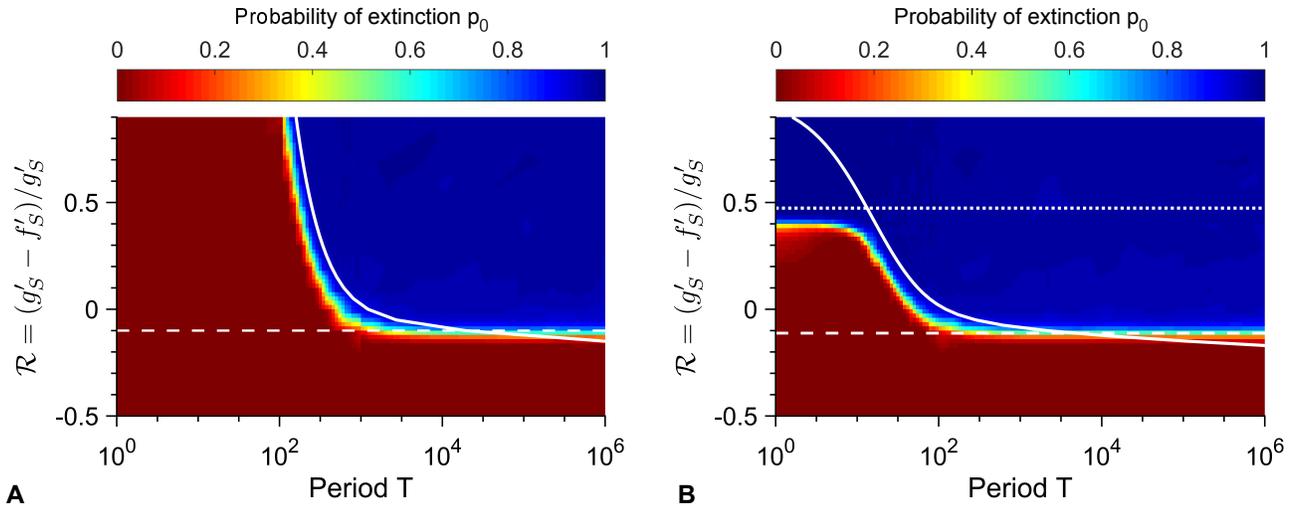}
	\vspace{0.2cm}
	\caption{{\bf Heatmaps of the extinction probability.} Extinction probability $p_0$ versus alternation period $T$ and  $\mathcal{R}=(g'_S-f'_S)/g'_S$ with biostatic (\textbf{A}) or biocidal (\textbf{B}) antimicrobial. Heatmap: simulation data, each point computed over $10^3$ realizations of simulation results, and linearly interpolated. Dashed white line: value of $\mathcal{R}$ such that $t_R^a=\tau_S$ (see main text). Solid white line: $T/2 = \tau_S$. Parameter values: $K=10^3$, $\mu_1=10^{-5}$, $\mu_2=10^{-3}$, $f_S=1$, $f_R=0.9$, $f_C=1$, $g_S=g_R=g_C=0.1$, and (\textbf{A}) $g'_S=0.1$ and variable $f'_S$ or (\textbf{B}) $f'_S=1$ and variable $g'_S$. Dotted line in \textbf{B}: $\mathcal{R}=(f_S-g_S)/(2f_S-g_S)$. All simulations start with 10 S microorganisms.}%
	\label{hm_f}%
\end{figure}

Fig.~\ref{hm_f} shows heatmaps of the probability $p_0$ that the microbial population goes extinct before resistance takes over, in the cases of biostatic and biocidal drugs, plotted versus the period of alternations $T$ and the non-dimensional variable $\mathcal{R}=(g'_S-f'_S)/g'_S$, which increases with antimicrobial concentration and is zero at the minimum inhibitory concentration (MIC). In both cases, two main regions are apparent, one with $p_0=0$ and one where $p_0$ is close to one. The transition between them is well described by the solid line $T/2=\tau_S$ such that the time spent with drug is equal to the extinction time $\tau_S$ of a population of sensitive microbes with drug, except for large periods, where the relevant transition occurs below the MIC ($\mathcal{R}<0$) and is given by $t_R^a=\tau_S$ (dashed line), consistently with our analytical predictions. 

The ratio $\mathcal{R}$ enables us to make a quantitative comparison between biostatic and biocidal drugs. 
Let us focus first on the transition $\tau_S=t_R^a$. Eq.~\ref{tj0} shows that the average time it takes for the sensitive microorganisms to spontaneously go extinct in the presence of antimicrobial can be written as $\tau_S(f'_S,g'_S)=\Phi(\mathcal{R})/g'_S$, where $\Phi$ is a non-dimensional function. Besides, the average fixation time of a R mutant in a population of S individuals can also be expressed as $t_R^a(f'_S,g'_S)=\Psi(\mathcal{R})/g'_S$, where $\Psi$ is a non-dimensional function. Thus, the transition $\tau_S=t_R^a$ will be the same for biostatic and biocidal drugs at a given value of $\mathcal{R}$. 
Conversely, the transition $\tau_S=T/2$, i.e. $\Phi(\mathcal{R})/g'_S=T/2$, depends on $g'_S$, and is thus different for biostatic and biocidal drugs at the same value of $\mathcal{R}$. Specifically, for a given value of $\mathcal{R}$, smaller periods $T$ will suffice to get extinction after the first addition of antimicrobial for a biocidal drug than for a biostatic drug, because $g'_S$ is increased by biocidal drugs, and hence $\tau_S$ is smaller in the biocidal case than in the biostatic case. This means that \textit{the parameter regime where treatment is efficient is larger for biocidal drugs than for biostatic drugs, as can be seen by comparing Fig.~\ref{hm_f}A and Fig.~\ref{hm_f}B}.  Significantly above the MIC, another difference is that \textit{biocidal drugs become efficient even for short periods $T/2\ll \tau_S$ if their concentration is large enough to have $g'_S>2f_S-g_S$, i.e. $\mathcal{R}>(f_S-g_S)/(2f_S-g_S)$} (see above, esp. Figs.~\ref{Per_pres_bioc}D-E). Numerical simulation results agree well with this prediction (dotted line on Fig.~\ref{hm_f}B).

\begin{figure}[htb]%
  	\centering
  	\includegraphics[width=0.95\textwidth]{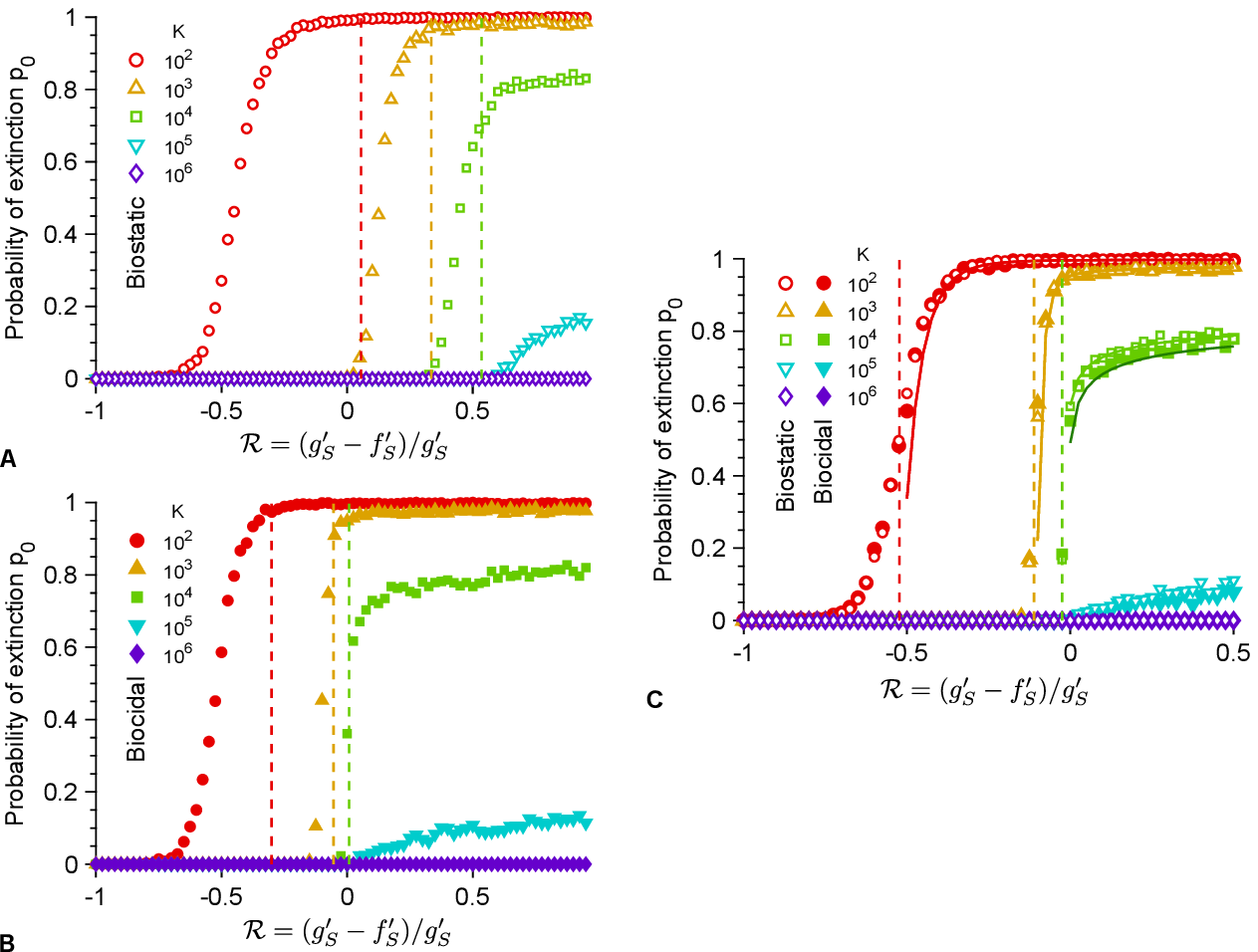}
  	\vspace{0.2cm}
  	\caption{{\bf Dependence of the extinction transition on population size and antimicrobial mode of action.} Extinction probability $p_0$ versus the ratio $\mathcal{R}=(g'_S-f'_S)/g'_S$ with biostatic or biocidal antimicrobial, for different carrying capacities $K$, either in the small-period regime, with $T=10^{2.5}$ (\textbf{A} and \textbf{B}) or in the large-period regime, with $T=10^5$ (\textbf{C}). Markers: simulation results, calculated over $10^3$ realizations. Vertical dashed lines: predicted extinction thresholds, i.e. values of $\mathcal{R}$ such that $T/2=\tau_S$ (\textbf{A} and \textbf{B}) or $t_R^a=\tau_S$ (\textbf{C}). Solid lines (\textbf{C}): Analytical estimates of $p_0$ from Eq.~\ref{p0_biostatic} (biostatic) or Eq.~\ref{p0_biocidal} (biocidal). For $K=10^2$ and $10^3$, the analytical predictions in the biostatic and biocidal case are confounded, while for $K=10^4$ we used two shades of green to show the slight difference (light: biostatic, dark: biocidal). Parameter values:  $\mu_1=10^{-5}$, $\mu_2=10^{-3}$, $f_S=1$, $f_R=0.9$, $f_C=1$, $g_S=g_R=g_C=0.1$, and $g'_S=0.1$ (biostatic) or $f'_S=1$ (biocidal). All simulations start with 10 S microorganisms.}%
  	\label{ic_f}%
  \end{figure}

\begin{figure}[h!]%
 	\centering
    \includegraphics[width=0.5\textwidth]{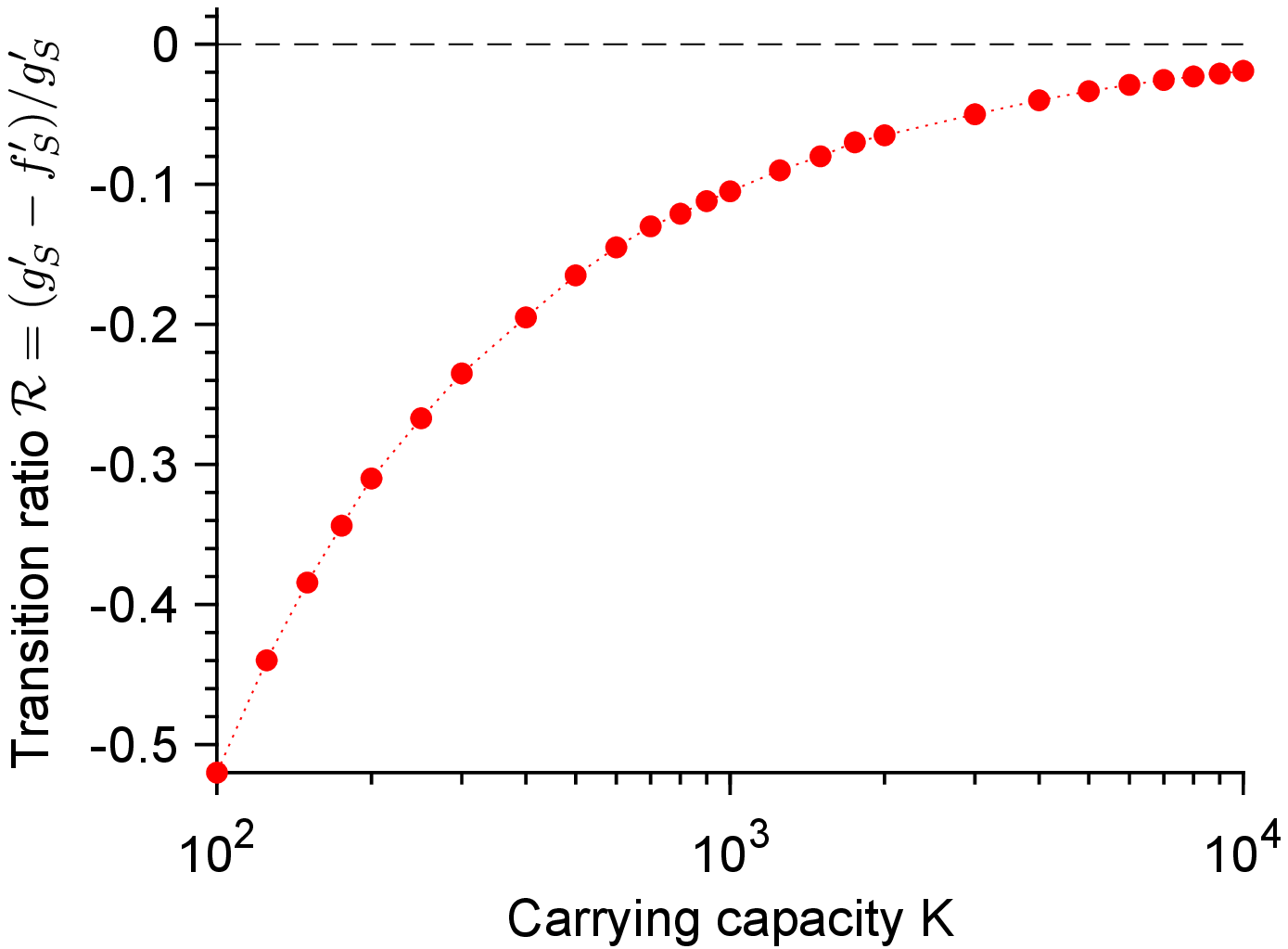}
 	\caption{{\bf Finite size effect on the extinction transition.} Value of the ratio $\mathcal{R}=(g'_S-f'_S)/g'_S$ such that $t_R^{a}=\tau_S$, plotted versus the carrying capacity $K$. This value of $\mathcal{R}$ marks the transition between large and small extinction probability $p_0$ when $T/2>\tau_S$ (see main text and Fig.~\ref{hm_f}). Red markers: numerical solutions of the equation $t_R^{a}=\tau_S$. Black dashed line: expected transition in the large population limit ($\mathcal{R}=0$, i.e. $f'_S=g'_S$). Parameter values: $\mu_1=10^{-5}$, $f_S=1$, $f_R=0.9$, $g_S=g_R=0.1$. Here, results are shown in the biostatic case, and $f'_S$ was varied, keeping $g'_S=0.1$, but the biocidal case yields the exact same results (see main text).}%
 	\label{ic_g}%
 \end{figure}

\begin{figure}[htb]%
  	\centering
  	\includegraphics[width=0.95\textwidth]{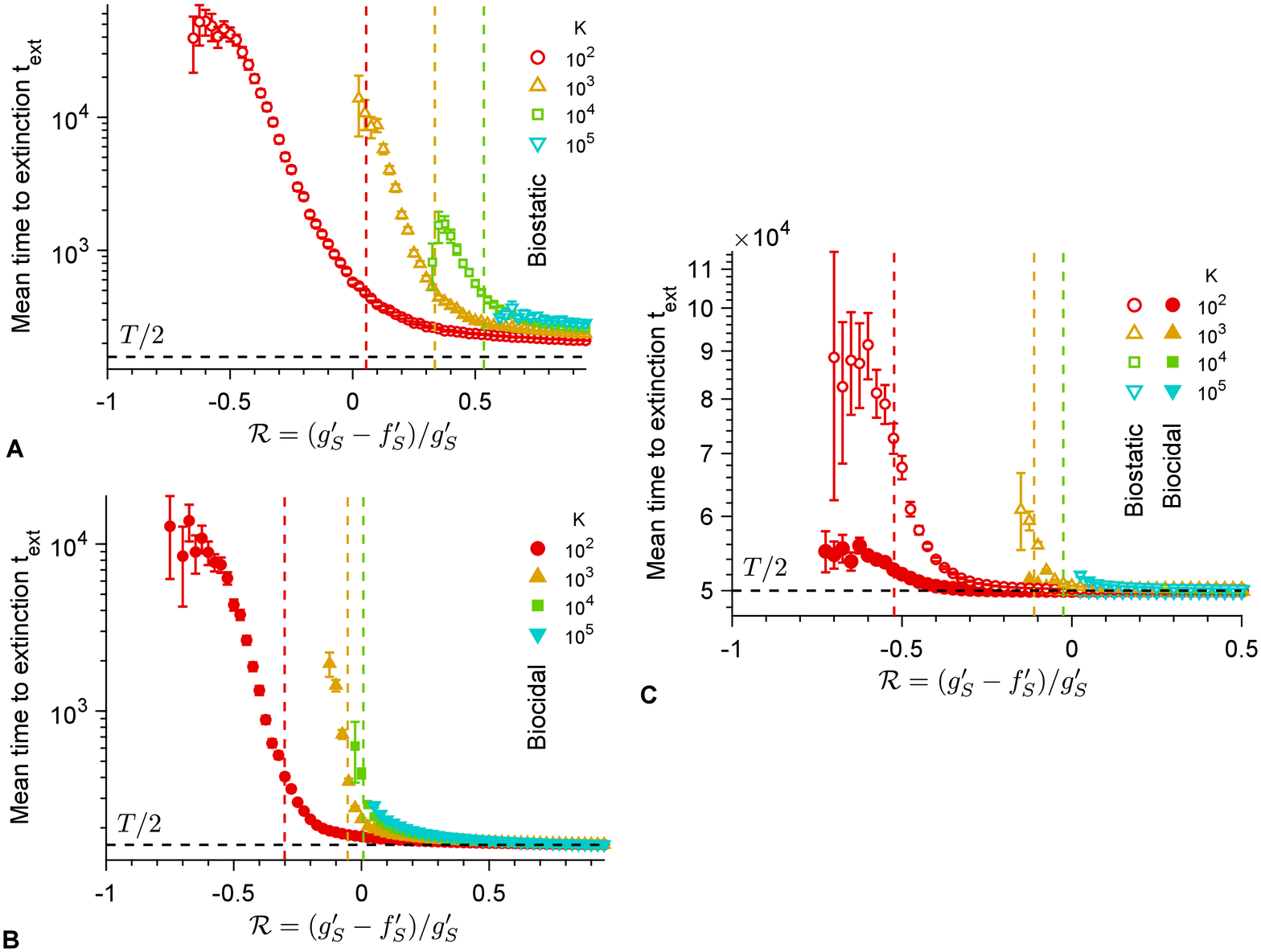}
  	\vspace{0.2cm}
  	\caption{{\bf Dependence of the average extinction time on population size and antimicrobial mode of action.} Average extinction time $t_{ext}$ versus the ratio $\mathcal{R}=(g'_S-f'_S)/g'_S$ with biostatic or biocidal antimicrobial, for different carrying capacities $K$, either in the small-period regime, with $T=10^{2.5}$ (\textbf{A} and \textbf{B}) or in the large-period regime, with $T=10^5$ (\textbf{C}). Markers: simulation results, calculated over the realizations ending in extinction of the population, if their number is at least 10, among $10^3$ realizations total per marker. Error bars: 95\% confidence intervals. Vertical dashed lines: predicted extinction thresholds, i.e. values of $\mathcal{R}$ such that $T/2=\tau_S$ (\textbf{A} and \textbf{B}) or $t_R^a=\tau_S$ (\textbf{C}). Horizontal dashed lines: $t_{ext}=T/2$. Parameter values (same as in Fig.~\ref{ic_f}):  $\mu_1=10^{-5}$, $\mu_2=10^{-3}$, $f_S=1$, $f_R=0.9$, $f_C=1$, $g_S=g_R=g_C=0.1$, and $g'_S=0.1$ (biostatic) or $f'_S=1$ (biocidal). All simulations start with 10 S microorganisms.}%
  	\label{text_fig}%
  \end{figure}

Importantly, \textit{the transition between large and small extinction probability when $\mathcal{R}$ (and thus the antimicrobial concentration) is varied strongly depends on population size, specifically on carrying capacity (Figs.~\ref{ic_f} and~\ref{ic_g}), and also depends on antimicrobial mode of action (Fig.~\ref{ic_f})}. For small periods where the relevant transition occurs for $T/2=\tau_S$, concentrations above the MIC ($\mathcal{R}>0$) can actually be necessary to get extinction because one period may not suffice to get extinction, and moreover, the extinction threshold value $\mathcal{R}$ is not the same for biostatic and biocidal antimicrobials (see above and Figs.~\ref{ic_f}A-B). Conversely, for large periods where the relevant transition occurs for $t^a_R=\tau_S$, and extinction occurs upon the first addition of drug, the extinction threshold is always below the MIC ($\mathcal{R}<0$) and it is the same for biostatic and biocidal antimicrobials (see above and Fig.~\ref{ic_f}C). In both cases, the larger the population, the larger the concentration required to get large extinction probabilities. For large periods (Fig.~\ref{ic_f}C), the transition occurs close to the MIC for large populations, but the smaller the population, the larger the discrepancy between the MIC and the actual transition, as predicted by our analytical estimate based on $t^a_R=\tau_S$ (see Fig.~\ref{ic_g}). This is because in small populations, stochastic extinctions of the population are quite fast at weakly sub-MIC antimicrobial. This is a form of inoculum effect, where the effective MIC depends on the size of the bacterial population~\cite{AlexanderPreprint}. In the large period regime (Fig.~\ref{ic_f}C), the extinction probability $p_0$ is well-predicted by Eqs.~\ref{p0_biostatic} and~\ref{p0_biocidal} for the $\mathcal{R}$ values such at most one R mutant can appear before the extinction of the population (as assumed in our calculation of $p_R^a$). In this regime, the extinction time is close to $T/2$ (see Fig.~\ref{text_fig}) as extinction is due to the first addition of antimicrobial, while for smaller $\mathcal{R}$ values, extinction occurs after multiple periods.
  
In Figs. \ref{ic_f}A-B, transitions between small and large values of $p_0$ in simulated data are observed for smaller threshold values of $\mathcal{R}$ than predicted by $T/2=\tau_S$ (this can also be seen in Fig.~\ref{hm_f}, where the solid white line is somewhat in the blue zone corresponding to large $p_0$). This is because we have employed the average extinction time $\tau_S$, while extinction is a stochastic process. Thus, even if $T/2<\tau_S$, upon each addition of antimicrobial, there is a nonzero probability that extinction actually occurs within the half-period. Denoting by $p$ the probability that a given extinction time is smaller than $T/2$, the population will  on average go extinct after $1/p$ periods, unless resistance fixes earlier. For instance, a population with carrying capacity $K=10^2$ submitted to alternations with $T=10^{2.5}$ is predicted to develop resistance before extinction if $\mathcal{R}<0.055$. However, for $\mathcal{R}=-0.1$, simulations yield a probability $p_0=0.99$ of extinction before resistance takes over (see Fig. \ref{ic_f}A). In this case, simulations yield $p=0.3$, implying that extinction typically occurs in $\sim\! 3$ periods, thus explaining the large value of $p_0$. More generally, the probability distribution function of the extinction time can depend on various parameters, which can impact the discrepancy between the predicted and observed transitions. A more precise calculation would involve this distribution. Note that the distribution of extinction times is known to be exponential for populations with a quasi-stationary state~\cite{Grimm04,Ovaskainen10}, but the present situation is more complex because there is no nonzero deterministic equilibrium population size below the MIC, and because the population size at the time when antimicrobial is added is far from the equilibrium value with antimicrobial. Nevertheless, our prediction based on the average extinction time $\tau_S$ yields the right transition shape (see Fig.~\ref{hm_f}) and the correct expectations for $T/2\gg\tau_S$ and  $T/2\ll\tau_S$.

\section*{Discussion}

\subsection*{Main results}

The evolution of antimicrobial resistance often occurs in variable environments, as antimicrobial is added and removed from a medium or given periodically to a patient, e.g. in a treatment by the oral route~\cite{Lin16,Levin-Reisman17}. Alternations of phases of absence and presence of antimicrobial induce a dramatic time variability of selection pressure on microorganisms, and can thus have a strong impact on resistance evolution. Using a general stochastic model which includes variations of both composition and size of the microbial population, we have shed light on the impact of periodic alternations of presence and absence of antimicrobial on the probability that resistance evolves \textit{de novo} and rescues a microbial population from extinction. The majority of previous studies of periodic antimicrobial treatments~\cite{Lipsitch97, Wahl00,Regoes04,SchulzzurWiesch10,Meredith15,Bauer17,Hansen17} neglect stochastic effects, while they can have a crucial evolutionary impact~\cite{Fisher07,Ewens79}, especially on population extinction~\cite{Coates18,AlexanderPreprint}. In addition, established microbial populations are structured, even within a single patient~\cite{VanMarle07}, and competition is local, which decreases their effective size, thus making stochasticity relevant. While a few previous studies did take stochasticity into account, some did not include logistic growth or compensation of the cost of resistance~\cite{NissenMeyer66}, while others made specific assumptions on treatments or epidemiology~\cite{AbelzurWiesch14,Ke15}, focused on numerical results with few analytical predictions~\cite{Wu14}, or assumed a constant population size~\cite{Marrec18}. The present model has the advantage of being quite general while fully accounting for stochasticity and finite-population effects.

We showed that fast alternations of presence and absence of antimicrobial are inefficient to eradicate the microbial population and strongly favor the establishment of resistance, unless the antimicrobial increases enough the death rate, which can occur for biocidal antimicrobials at high concentration~\cite{Levin10,Khan15}. The corresponding criterion on the death rate $g'_S$ of sensitive microorganisms with biocidal antimicrobial, namely $g'_S>2f_S-g_S$, is generally more stringent than simply requiring drug concentrations to be above the MIC during the phases with biocidal antimicrobial, namely $g'_S>f_S$. Indeed, the population can re-grow without antimicrobial: in this regime, extinction occurs over multiple periods, and involves decaying oscillations. Conversely, for biostatic antimicrobials, as well as for biocidal ones at smaller concentrations, extinction has to occur within a single phase with antimicrobial, and thus the half-period $T/2$ has to be longer than the average extinction time $\tau_S$, which we fully expressed analytically. Importantly, shorter periods suffice for biocidal antimicrobials compared to biostatic ones in order to drive a population to extinction upon the first addition of antimicrobial, at the same value of $\mathcal{R}=(g'_S-f'_S)/g'_S$. Hence, the parameter regime where treatment is efficient is larger for biocidal drugs than for biostatic drugs. If $T/2>\tau_S$, the microbial population goes extinct upon the first addition of antimicrobial, unless it is rescued by resistance. We obtained an analytical expression for the probability $p_0$ that the population is eradicated upon the first addition of antimicrobial, assuming rare mutations. Note that with realistic bacterial mutation probabilities, namely $\mu_1\sim 10^{-10}$~\cite{Wielgoss11}, the rare mutation regime remains relevant even for quite large populations. Moreover, real microbial populations are generally structured, which reduces their effective population size. Rescue by resistance can happen either if resistant mutants preexist upon the addition of antimicrobial, or if they appear after antimicrobial is added to the environment, during the decay of the population. Importantly, the latter case is fully prevented by perfect biostatic antimicrobials that completely stop division of sensitive microorganisms. This sheds light on the respective merits of different antimicrobial modes of action. Finally, we showed that due to stochastic extinctions, sub-MIC concentrations of antimicrobials can suffice to yield extinction of the population, and we fully quantified this effect and its dependence on population size. Throughout, all of our analytical predictions were tested by numerical simulations, and the latter also allowed us to explore cases beyond the rare mutation regime, where resistance occurs more frequently.

This work opens many possible theoretical extensions. In particular, it will be very interesting to include effects such as antibiotic tolerance, which tend to precede resistance under intermittent antibiotic exposure~\cite{Levin-Reisman17}, as well as to consider the possibility of concentrations above the mutant prevention concentration, such that resistant microbes are also affected by the drug~\cite{Levin-Reisman17,Bauer17}. Another exciting extension would be to incorporate spatial structure~\cite{Bitbol14,Nahum15,Cooper15} and environment heterogeneity, in particular drug concentration gradients. Indeed, static gradients can strongly accelerate resistance evolution~\cite{Zhang11,Greulich12,Hermsen12,Baym16}, and one may ask how this effect combines with the temporal alternation-driven one investigated here. Besides, it would be interesting to explicitly model horizontal gene transfer of resistance mutations, to include realistic pharmacodynamics and pharmacokinetics~\cite{Regoes04}, and also to compare the impact of periodic alternations to that of random switches of the environment~\cite{Mustonen08,Rivoire11,Melbinger15,Desponds16,Hufton16,Wienand17,Meyer18,Danino18}. Other effects such as single-cell physiological properties~\cite{Lin16}, phenotypic delay~\cite{CarballoPreprint} or density dependence of drug efficacy~\cite{HallinenPreprint} can further enrich the response of microbial populations to variable concentrations of antimicrobials.

\subsection*{Practical relevance}

Our results have consequences for actual experimental and clinical situations. First, several of our predictions can be tested experimentally in controlled setups such as that presented in Ref.~\cite{Lin16}. This would allow for an experimental test of the transition of extinction probability between the short-period and the long-period regimes, and of the predicted values of this extinction probability for large periods in the rare mutation regime. Second, the situation where the phases of absence and presence of antimicrobial have similar durations, which we considered here, is unfortunately clinically realistic. Indeed, a goal in treatment design is that the serum concentration of antimicrobial exceeds the MIC for at least 40 to 50\% of the time~\cite{Jacobs01}. Because bacteria divide on a timescale of about an hour in exponential growth phase, and because antimicrobial is often taken every 8 to 12 hours in treatments by the oral route, the alternation period lasts for a few generations in treatments: this is the same order of magnitude as the transition we found between the short-period and long-period regimes, meaning that this transition is relevant in clinical cases. Note that while this transition timescale depends on the death and birth rates of sensitive microbes in the presence of antimicrobial (see Eq.~\ref{tj0}), and therefore on antimicrobial concentration, it does not depend on the value of the mutation rate or on the initial population size (as long as the half-period is longer than the initial population growth timescale, see Supporting Information, section~\ref{tr_SI}), and it depends only weakly on the carrying capacity, e.g. logarithmically in the perfect biostatic case (see Eq.~\ref{t_j0_a}). Given the relevance of this transition between the short-period and the long-period regimes, it would be very interesting to conduct precise measurements of both division rates and death rates~\cite{Frenoy18} in actual infections in order to determine the relevant regime in each case. This is all the more important that in the short-period regime, we showed that only large concentrations of biocidal antimicrobials are efficient, while other antimicrobials systematically lead to the \textit{de novo} evolution of resistance before eradication of the microbial population. This constitutes a striking argument in favor of the development of extended-release antimicrobial formulations~\cite{Gao11}.  Conversely, a broader spectrum of modes of action can be successful for longer periods of alternation of drug absence and presence. 

Despite the fact that only biocidal antimicrobials at high concentration are efficient for short alternation periods of absence and presence of drug, and the fact that the parameter regime where treatment is efficient is larger for biocidal drugs than for biostatic drugs, biostatic antimicrobials that fully stop division of sensitive microorganisms have a distinct advantage over drugs with other modes of action. Indeed, they prevent the emergence of resistant mutants when drug is present, which is all the more important that such resistant mutants are immediately selected for by the antimicrobial and are thus quite likely to rescue the microbial population and to lead to the fixation of resistance. This argues in favor of combination therapies involving a biostatic and a biocidal antimicrobial. Note however that the combined drugs need to be chosen carefully, because some of them have antagonistic interactions~\cite{Bollenbach09}, depending on their mode of action.

\section*{Acknowledgments}
The authors thank Claude Loverdo for inspiring discussions. LM acknowledges
funding by a graduate fellowship from EDPIF.


\newpage

\beginsupplement


\begin{center}
\huge{\textbf{Supporting Information}}
\end{center}

\vspace{1cm}

\tableofcontents

\newpage

\section{Population with a single type of microorganisms}

\subsection{Master equation}  

Let us first consider the simple case of a microbial population with a carrying capacity $K$ comprising a single type of microorganisms. These microorganisms have a fitness and a death rate denoted by $f$ and $g$, respectively. Let $j$ be the number of individuals in the population at time $t$, satisfying $0 \leq j \leq K$. The master equation describing the evolution of this population reads for all $j$:
\begin{equation}
\frac{dP_j(t)}{d t}=f\left(1-\frac{j-1}{K}\right)(j-1)P_{j-1}(t)+g(j+1)P_{j+1}(t)-\left( f\left(1-\frac{j}{K}\right)+g\right)jP_j(t)\mbox{ }.
\label{mast_eq}
\end{equation}
Indeed, recall that $f(1-j/K)$ is the division rate in the logistic model. We can write this system of equations as ${\bf \dot{P}}={\bf RP}$, where ${\bf R}$ is the transition rate matrix:
\begin{equation}
\frac{d}{dt}\begin{pmatrix}
P_0 \\
P_1 \\
P_2 \\
\vdots \\
P_K
\end{pmatrix}=
\begin{pmatrix}
0 & g & 0 & \cdots & 0\\
0 & -g-f(1-1/K) & 2g & (0) & \vdots\\
0 & f(1-1/K) & -2g-2f(1-2/K) & \ddots & 0\\
\vdots & (0) & \ddots & \ddots & Kg \\
0 & \cdots & 0 & f(1-(K-1)/K)(K-1) & -Kg
\end{pmatrix}
\begin{pmatrix}
P_0 \\
P_1 \\
P_2 \\
\vdots \\
P_K
\end{pmatrix}\mbox{ }.
\label{mast_eq_vect}
\end{equation}
This Markov chain has a single absorbing state, namely $j=0$,  which corresponds to the extinction of the microbial population.

\subsection{Average spontaneous extinction time}
\label{SI_mfpt}

Let us study the average time it takes for the population to spontaneously go extinct, i.e. the mean first-passage time $\tau_S(j_0)$ to the absorbing state $j=0$, starting from $j_0$ microorganisms at $t=0$. It can be expressed using the inverse of the reduced transition rate matrix ${\bf \widetilde{R}}$, which is identical to $\bf{R}$ except that the row and the column corresponding to the absorbing state $j=0$ are removed~\cite{Marrec18,Sekimoto10}:
\begin{equation}
\tau_S(j_0)=\mathbb{E}[\widehat{\tau}_{FP} \, | \, j_0]=-\sum_{i=1}^K ({\bf \widetilde{R}}^{-1})_{i \, j_0}\mbox{ }.
\label{MFPT}
\end{equation}
Note that more generally, all the moments of the first-passage time can be obtained using the reduced transition rate matrix ${\bf \widetilde{R}}$:
\begin{equation}
\mathbb{E}[\widehat{\tau}_{FP}^n \, | \, j_0]=n!(-1)^n\sum_{i=1}^K ({\bf \widetilde{R}}^{-n})_{i \, j_0}\mbox{ }.
\label{n_moment_FPT}
\end{equation}
Here, the elements of the inverse of the reduced transition matrix read for all $1\leq j\leq K$,
\begin{equation}
({\bf \widetilde{R}}^{-1})_{i \, j} = \begin{cases} \displaystyle -\sum_{k=0}^{i-1}\left(\frac{f}{g}\right)^{i-k-1}\frac{K^{k+1-i}(K-k-1)!}{i\,g\,(K-i)!} & \mbox{if } i \leq j\,,\\
\displaystyle -\sum_{k=0}^{j-1}\left(\frac{f}{g}\right)^{i-k-1}\frac{K^{k+1-i}(K-k-1)!}{i\,g\,(K-i)!}&\mbox{if } i>j\,.
 \end{cases} 
\label{Red_trans_mat}
\end{equation}
Substituting Eq. \ref{Red_trans_mat} in Eq. \ref{MFPT} yields
\begin{equation}
\tau_S(j_0)=\frac{1}{g}\left[\sum_{i=1}^{j_0} \sum_{k=0}^{i-1} \left(\frac{f}{g}\right)^{i-k-1}\frac{K^{k+1-i}(K-k-1)!}{i \, (K-i)!}+\sum_{i=j_0+1}^{K} \sum_{k=0}^{j_0-1} \left(\frac{f}{g}\right)^{i-k-1}\frac{K^{k+1-i}(K-k-1)!}{i \, (K-i)!}\right]\,.
\label{tj0}
\end{equation}

If $f=0$, e.g. in the presence of a biostatic antimicrobial that perfectly prevents all microorganisms from growing, Eq. \ref{tj0} simplifies to:
\begin{equation}
\tau_S(j_0)=\frac{1}{g}\sum_{i=1}^{j_0} \frac{1}{i} \mbox{ }.
\label{t_j0_a}
\end{equation}
Note the formal analogy between Eq. \ref{t_j0_a} and the unconditional fixation time with biostatic antimicrobial ($f=0$) in the Moran process, which corresponds to the extinction of the sensitive microbes in a population of fixed size~\cite{Marrec18}. Both situations involve the extinction of microorganisms that do not grow. Formally, the master equation of a Moran process describing a microbial population of fixed size $N$ with two types of individuals A and B whose respective fitnesses are $f_A=0$ and $f_B=1$, reads:
\begin{equation}
\frac{d P_l(t)}{d t}=\frac{l+1}{N}P_{l+1}(t)-\frac{l}{N}P_l(t)\mbox{ },
\end{equation}
where $l$ denotes the number of A individuals. The master equation for a logistic growth of a population with a single type of individuals (see Eq. \ref{mast_eq}) with $f=0$ is equivalent under the transformation $1/N \leftarrow g$.

\begin{figure}[htb]
	\centering
	\includegraphics[width=\textwidth]{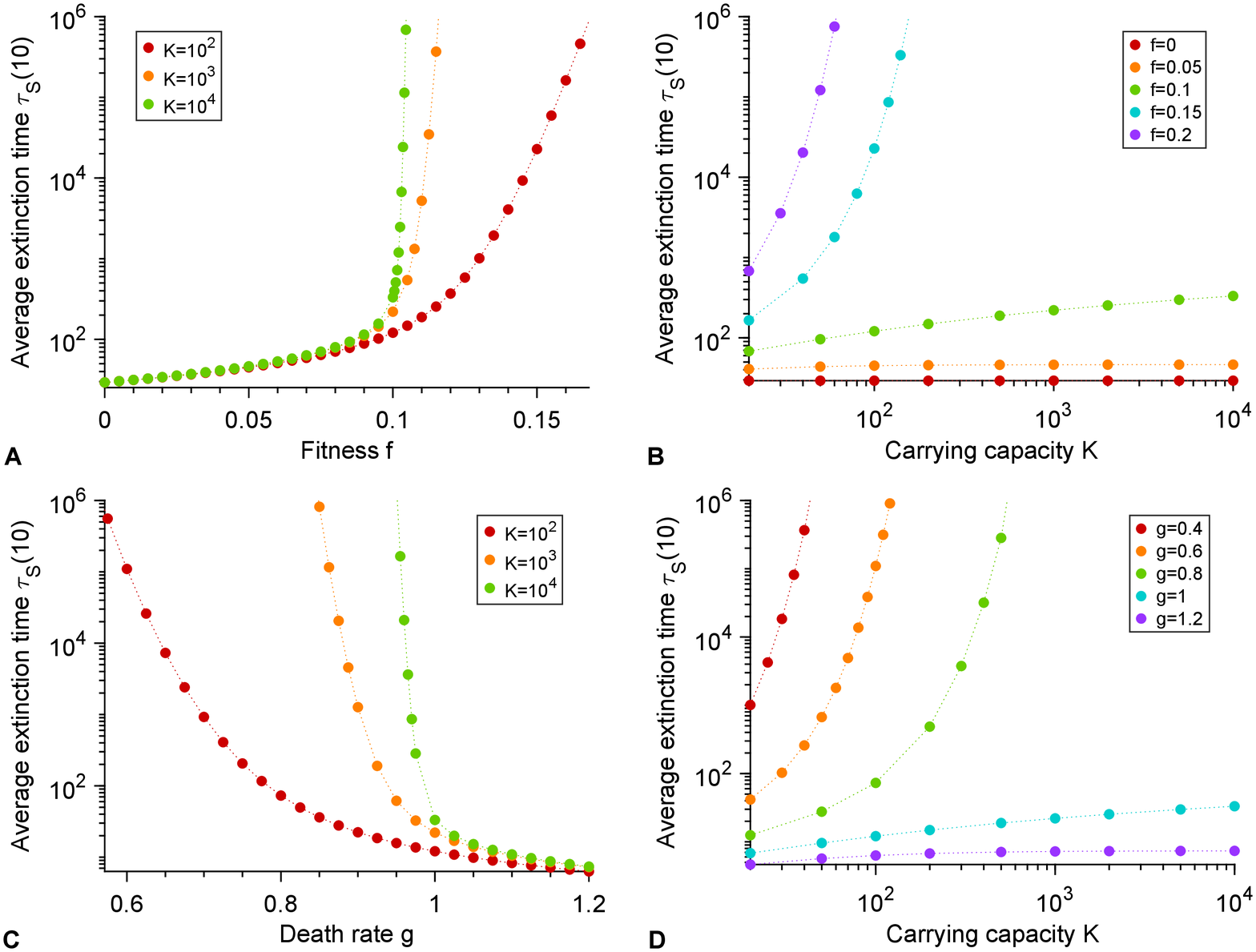}
	\vspace{0.2cm}
	\caption{{\bf Average spontaneous extinction time of the microbial population.} \textbf{A:} Mean first-passage time $\tau_S(10)$ to the absorbing state $j=0$, i.e. average extinction time, starting from $j_0=10$ microorganisms, as a function of the fitness $f$ for different carrying capacities $K$, with $g=0.1$. \textbf{B:} Average extinction time $\tau_S(10)$ as a function of the carrying capacity $K$ for different fitnesses $f$, with $g=0.1$. \textbf{C:} Average extinction time $\tau_S(10)$ as a function of the death rate $g$ for different carrying capacities $K$, with $f=1$. \textbf{D:} Average extinction time $\tau_S(10)$ as a function of the carrying capacity $K$ for different death rates $g$, with $f=1$.  }%
	\label{t_1}%
\end{figure}

Fig. \ref{t_1} shows how $\tau_S(10)$ depends on the death rate $g$ and the carrying capacity $K$. In particular, it shows that when $g<f$, average extinction times become very long for large values of $K$, while they are short for all $K$ when $g>f$. In a deterministic description (valid for very large population sizes), $g=f$ indeed corresponds to the transition between a population that decays exponentially and a population that reaches a steady state size. For finite-sized populations, stochasticity makes this transition smoother.

\newpage

\subsection{Initial growth of the population}

\subsubsection{Deterministic approximation and rise time}

\label{tr_SI}
\begin{figure}[h!]
\centering
\includegraphics[width=\textwidth]{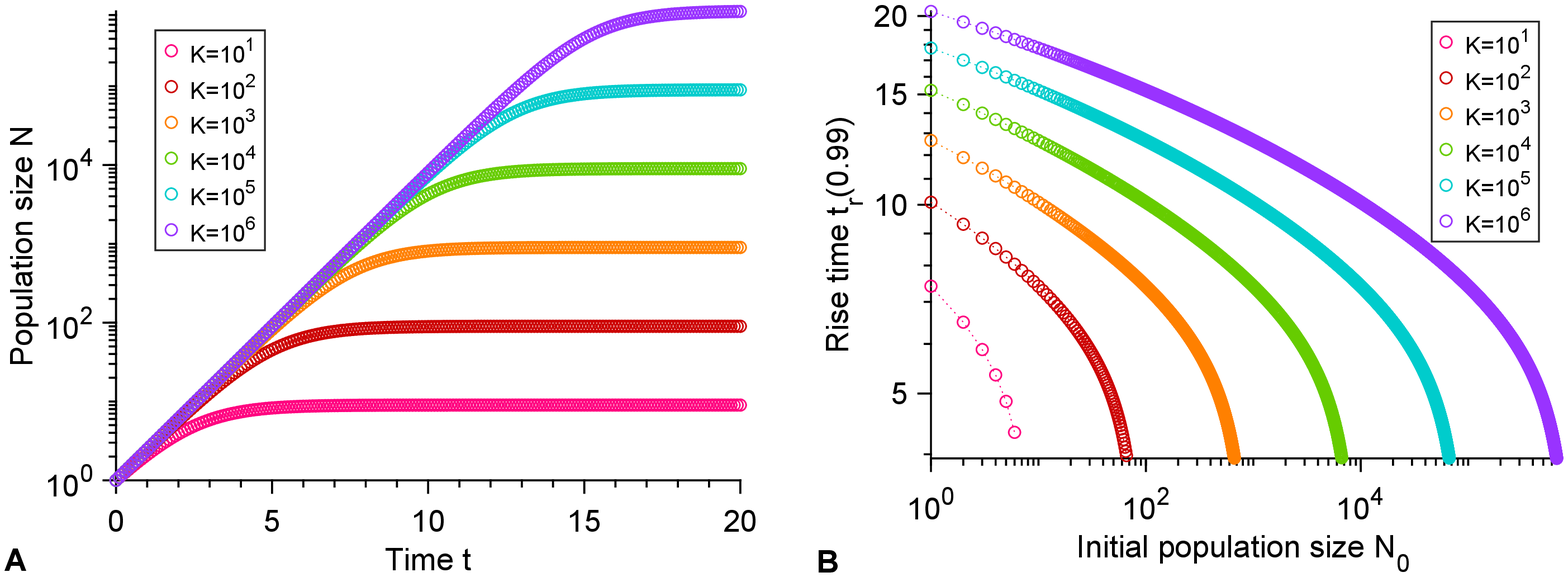}
	\caption{{\bf Deterministic evolution of the population size and rise time.} \textbf{A:} Population size $N$ as function of time $t$ for different carrying capacities $K$. \textbf{B:} rise time $t_r(0.99)$ as function of the initial number of individuals $N_0$ for different carrying capacities $K$. Results are obtained from Eqs. \ref{N_t} and \ref{t_r}. Parameter values: $f=1$ and $g=0.1$. }%
	\label{N_t_and_t_r}%
\end{figure}

 In the deterministic regime, for a population with only one type of microorganisms and a carrying capacity $K$, the number $N$ of individuals at time $t$ follows the logistic ordinary differential equation:
\begin{equation}
\frac{dN(t)}{dt}=N(t)\left[f\left(1-\frac{N(t)}{K}\right)-g\right] \,,
\label{edo}
\end{equation}
where $f$ represents fitness and $g$ death rate. For $f\neq g$, the solution reads:
\begin{equation}
N(t)=\frac{K \, N_0 \, e^{(f-g)t} \, (1-g/f)}{K \, (1-g/f)+N_0 \, (e^{(f-g)t}-1)} \,,
\label{N_t}
\end{equation}
where $N_0=N(0)$ is the initial number of individuals in the population. Note that we recover the usual law of logistic population growth for $f>0$ and $g=0$ (or for $f>g$ by setting $f\leftarrow f-g$):
\begin{equation}
N(t)=\frac{K \, N_0 \, e^{f \, t}}{K+N_0 \, (e^{f \, t}-1)} \,.
\end{equation}
For $f>g$, the long-time limit of Eq.~\ref{N_t} is $K(1-g/f)$. This equilibrium population size can also be found as the steady-state solution of Eq.~\ref{edo}, and corresponds to the birth and death rates being equal. The rise time $t_r(\alpha)$, at which a fraction $\alpha$ of this equilibrium population size is reached, is given by:
\begin{equation}
t_r(\alpha)=\frac{1}{f-g}\ln\left(\frac{\alpha K (1-g/f)-\alpha N_0}{(1-\alpha)N_0}\right)\,.
\label{t_r}
\end{equation} 
Hence, the initial growth of the population is governed by the timescale $1/(f-g)$, and features a weaker dependence on carrying capacity $K$ and initial population size $N_0$, as illustrated by Fig.~\ref{N_t_and_t_r}.

\subsubsection{Probability of rapid initial extinction}

A microbial population starting with few individuals may go extinct quickly due to stochastic fluctuations, before reaching a substantial fraction of its equilibrium size $K(1-g/f)$. Formally integrating the master equation ${\bf \dot{P}}={\bf RP}$ with the initial condition $j=j_0$ allows to express the probability $P_0(t)$ that a population starting from $j_0$ microorganisms at $t = 0$ is extinct at time $t$:
\begin{equation}
P_0(t) = (e^{{\bf R}t})_{0 j_0} \,.
\label{quick_ext}
\end{equation}

\begin{figure}[h!]
	\begin{center}
	\includegraphics[width=\textwidth]{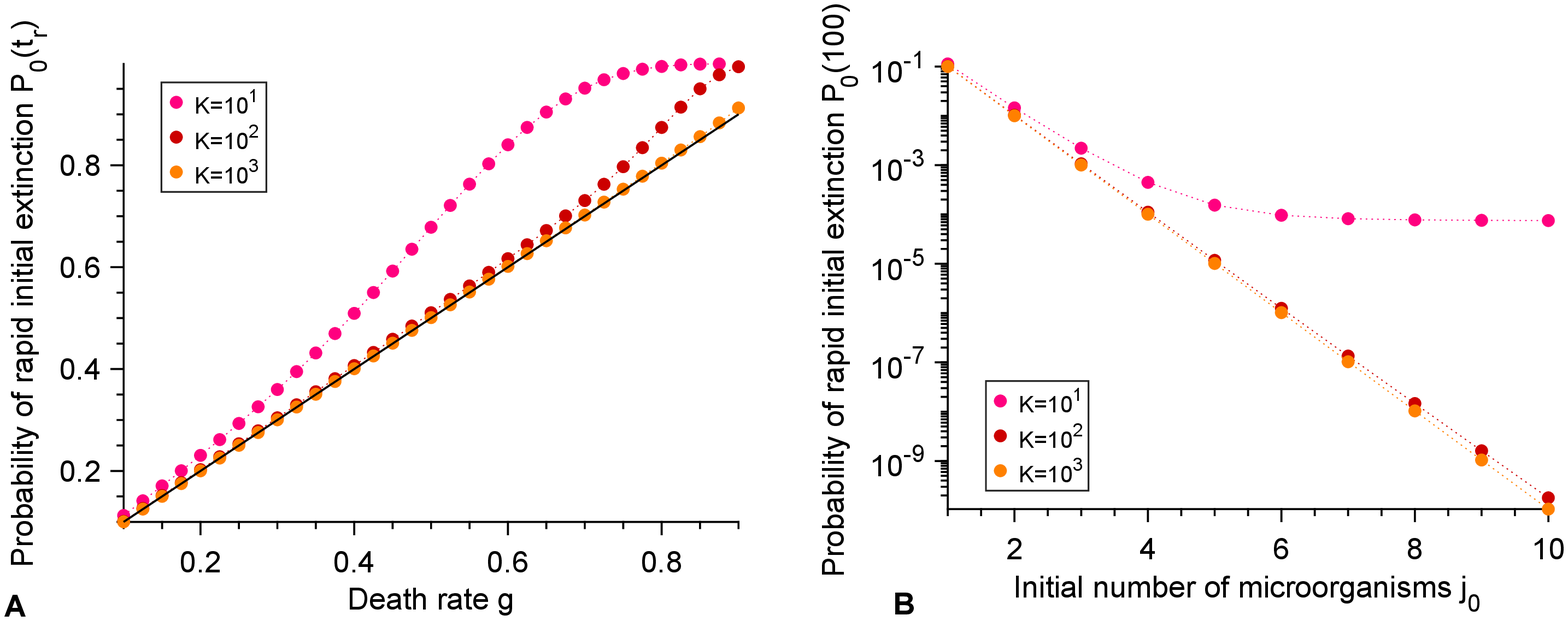}
		\caption{{\bf Rapid initial extinction}. \textbf{A:} Probability $P_0(t_r)$ that extinction occurs before the rise time $t_r(0.99)$ (see Eq.~\ref{t_r}), when starting from a single microorganism, $j_0=1$, as function of the death rate $g$ with $f=1$ for different carrying capacities $K$. Results come from a numerical computation of Eq. \ref{quick_ext}. Solid black line: $g/f$. \textbf{B:} Probability of rapid initial extinction $P_0(100)$ as a function of the initial number of microorganisms $j_0$, for different carrying capacities $K$. Data points correspond to numerical computations of Eq. \ref{quick_ext}. Parameter values: $f_S=1$ and $g_S=0.1$. Time $t=100$ was chosen to evaluate $P_0$ because it is larger than typical rise times for the parameter values considered (see Fig.~\ref{N_t_and_t_r}), but not too long, and thus captures rapid initial extinctions but not long-term ones (see Fig.~\ref{t_1}). }
		\label{quick_ext_plot}
	\end{center}
\end{figure}

Fig. \ref{quick_ext_plot} shows the probability $P_0(t_r)$ that the microbial population goes extinct before the rise time $t_r$ versus $g$ for $f=1$. We notice that $P_0(t_r)\sim g/f$ for small $g$ and/or large $K$. This result can be proved analytically by assuming that the number of individuals is very small compared to the carrying capacity $K$ and thus grows exponentially, which is relevant when rapid initial extinctions occur. One can then neglect the impact of the carrying capacity $K$ in the master equation Eq.~\ref{mast_eq}, yielding:
\begin{equation}
\frac{d P_j(t)}{d t}=f(j-1)P_{j-1}(t)+g(j+1)P_{j+1}(t)-\left( f+g\right)jP_j(t)\mbox{ }.
\end{equation}
The solution of this master equation is given by~\cite{Kendall48}:
\begin{equation}
P_j(t)=e^{(f-g)t}\left(\frac{1-g/f}{e^{(f-g)t}-g/f}\right)^2\left(\frac{e^{(f-g)t}-1}{e^{(f-g)t}-g/f}\right)^{j-1} \mbox{ }. 
\end{equation}
In particular, we thus obtain:
\begin{equation}
P_0(t)=\frac{g}{f}\left( \frac{e^{(f-g)t}-1}{e^{(f-g)t}-g/f} \right) \underset{t \rightarrow +\infty}{\rightarrow} \frac{g}{f} \mbox{  if }f>g \mbox{ }. 
\end{equation}

\clearpage

\section{Supplementary results on extinction probabilities and extinction and fixation times}

\subsection{Perfect biostatic antimicrobial}

\clearpage
\subsection{Biocidal antimicrobial}

\begin{figure}[h!]%
	\centering
	\includegraphics[width=\textwidth]{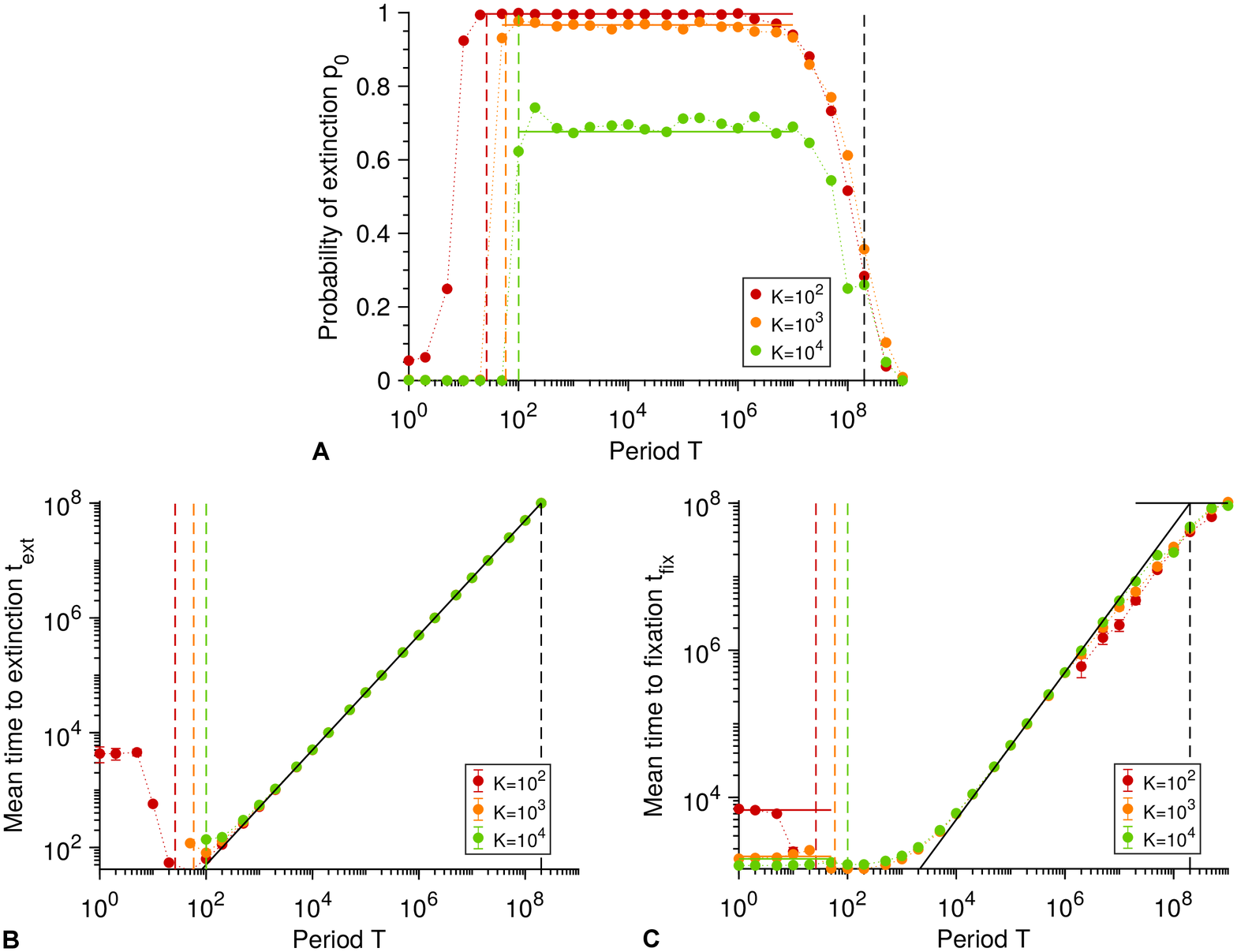}
	\vspace{0.2cm}
	\caption{{\bf Periodic presence of a biocidal antimicrobial above the MIC, including long periods.} \textbf{A:} Probability $p_0$ that the microbial population goes extinct before resistance gets established versus alternation period $T$, for various carrying capacities $K$. Markers: simulation results, with probabilities estimated over $10^2 - 10^3$ realizations. Horizontal solid lines: analytical predictions from Eq.~\ref{p0_biocidal}. \textbf{B:} Average time $t_{ext}$ to extinction versus alternation period $T$ for various carrying capacities $K$. Data shown if extinction occurred in at least 10 realizations. \textbf{C:} Average time  $t_{fix}$ to fixation of the C microorganisms versus alternation period $T$ for various carrying capacities $K$. Data shown if resistance took over in at least 10 realizations. Horizontal solid colored lines: analytical predictions for very small $T$, using the self-averaged death rate $\tilde{g}_S$ (see below). Horizontal solid black line: average spontaneous valley crossing time $\tau_V=(f_S-f_R)/(\mu_1\mu_2g_S)$ (see main text). In panels \textbf{B} and \textbf{C}, markers are averages over $10^2 - 10^3$ simulation realizations, error bars (often smaller than markers) represent $95\%$ confidence intervals, and the oblique black line corresponds to $T/2$. In all panels, colored dashed lines correspond to $T/2=\tau_S$, while black dashed lines correspond to $T/2=\tau_V$. Parameter values: $f_S=1$, $f_R=0.9$, $f_C=1$, $g_S=0.1$ without antimicrobial, $g'_S=1.1$ with antimicrobial, $g_R=g_C=0.1$, $\mu_1=10^{-5}$ and $\mu_2=10^{-3}$. All simulations start with 10 S microorganisms. }%
	\label{Per_pres_bioc_vc}%
\end{figure}

Here, in the limit of very fast alternations, we expect an effective averaging of death rates, with $\tilde{g}_S=0.6$ for S microorganisms. Then, an R mutant that will fix in the population appears after an average time $\tilde{t}_R^a=1/(\tilde{N}\mu_1\tilde{g}_S \tilde{p}_{SR})$ where $\tilde{N}\mu_1\tilde{g}_S$ represents the total mutation rate in the population, with $\tilde{N}=K(1-\tilde{g}_S/f_S)$ the equilibrium population size, and where $\tilde{p}_{SR}=[1-f_Sg_R/(f_R\tilde{g}_S)]/[1-(f_Sg_R/(f_R\tilde{g}_S))^{\tilde{N}}]$ is the probability that a single R mutant fixes in a population of $\tilde{N}$ microorganisms where all other microorganisms are S. Subsequently, C mutants will appear and fix, thus leading to the full evolution of resistance by the population. The corresponding average total time $t_{fix}$ of resistance evolution~\cite{Marrec18} agrees well with simulation results for $T/2 \ll \tau_S$ (see Fig.~\ref{Per_pres_bioc_vc}C).

\subsection{Population size dependence of the extinction transition}

\clearpage

\subsection{Dependence of the extinction time on population size and antimicrobial mode of action}

\clearpage

\section{Rescue by resistance}

\subsection{Number of resistant mutants when antimicrobial is added: $p_R^c(i)$}
\label{pRc_SI}

Let $p_R^c(i)$ be the probability that exactly $i$ R microorganisms are present when antimicrobial is added, provided that a lineage of R mutants then exists. It can be calculated in the framework of the Moran model, provided that the population size is stable around $N=K(1-g_S/f_S)$ before antimicrobial is added, which is correct for $T/2\gg t_r$, where $t_r$ is the rise time (see section~\ref{tr_SI}). Specifically, $p_R^c(i)$ can be expressed as a ratio of the sojourn time in state $i$ to the total lifetime of the lineage in the absence of antimicrobial: 
\begin{equation}
p_R^c(i)=\frac{\tau_{R,i}^d}{\tau_R^d}\,,
\end{equation} 
where $\tau_R^d$ is the average lifetime without antimicrobial of the lineage of a resistant mutant, assuming that it is destined for extinction, and  $\tau_{R,i}^d$ is the average time this lineage spends with exactly $i$ R individuals before going extinct. They satisfy $\tau_{R}^d=\sum_{i=1}^{N-1}\tau_{R,i}^d$. Note that we consider lineages destined for extinction in the absence of antimicrobial, because we focus on timescales much shorter than the spontaneous valley crossing time. In fact, in this regime, considering unconditional times yields nearly identical values for $p_R^c(i)$.

Employing the master equation ${\bf \dot{P}}={\bf RP}$ that describes the time evolution of the number of R mutants within the Moran model~\cite{Ewens79,Marrec18}, where ${\bf R}$ is the transition rate matrix, we obtain 
\begin{equation} 
\tau_{R,i}^d=\frac{\pi_i}{\pi_1}\int_{0}^{\infty}P_i(t)dt=-\frac{\pi_i}{\pi_1}({\bf \tilde{R}}^{-1})_{i \, 1}\,,
\end{equation} 
where $\pi_i$ is the probability that the R mutants go extinct, starting from $i$ R mutants~\cite{Ewens79,Marrec18}, while ${\bf \tilde{R}}$ is the reduced transition rate matrix, which is identical to the transition rate matrix ${\bf{R}}$, except that the rows and the columns corresponding to the absorbing states $i=0$ and $i=N$ are removed~\cite{Marrec18}. Here, we take $N=K(1-g_S/f_S)$, which corresponds to the deterministic equilibrium population size. Finally, we obtain
\begin{equation}  
p_R^c(i)=\frac{\pi_i({\bf \tilde{R}}^{-1})_{i \, 1}}{\sum_{k=1}^{N-1}\pi_k({\bf \tilde{R}}^{-1})_{k \, 1}}\,. 
\label{pRci}
\end{equation} 

\subsection{Probability of fast extinction of the resistant mutants: $p_R^e(i)$}
\label{pRe_SI}

Let us consider the beginning of the first phase with antimicrobial, and take as our origin of time $t=0$ the beginning of the phase with antimicrobial. Here, we consider the general case of an antimicrobial that may modify both the division rate and the death rate of sensitive microorganisms. Provided that some resistant microorganisms are present at $t=0$, how likely is it that they will undergo a rapid stochastic extinction and not rescue the microbial population and lead to the establishment of resistance? Denoting by $i>0$ the number of resistant microorganisms at $t=0$, let us estimate the probability $p_R^e(i)$ that the lineage of R mutants then quickly goes extinct. As explained in the main text, we approximate the reproduction rate of the R microorganisms by 
\begin{equation}
f_R(t)=f_R\left(1-\frac{S(t)+R(t)}{K}\right)\approx f_R\left(1-\frac{S(t)}{K}\right)\,,
\label{det1}
\end{equation}
where $S(t)$ and $R(t)$ are the numbers of S and R individuals at time $t$. This is appropriate because early extinctions of R mutants tend to happen shortly after the addition of antimicrobials, when $S(t)\gg R(t)$. Thus motivated, we further employ the deterministic approximation to describe the decreasing number $S(t)$ of S microorganisms:
	\begin{equation}
	S(t)=\frac{K(1-g'_S/f'_S)S_0e^{(f'_S-g'_S)t}}{K(1-g'_S/f'_S)+S_0(e^{(f'_S-g'_S)t}-1)}\mbox{ },
	\label{detg}
	\end{equation}
	where $S_0=K(1-g_S/f_S)$ is the number of sensitive microorganisms when antimicrobial is added. Note that if $f'_S=0$ and $g'_S=g_S$, i.e. in the perfect biostatic case, we obtain
\begin{equation} 
S(t)=K\left(1-\frac{g_S}{f_S}\right)e^{-g_St}\,,
\label{det2}
\end{equation}
for the decay of the number of S microorganisms with antimicrobial. However, we retain a stochastic description for the rare R mutants, and employ the probability generating function
\begin{equation}
\phi_{i}(z,t)=\sum_{j=0}^{\infty}z^jP(j,t|i,0)\,,
\label{PGF}
\end{equation}
where $i$ is the initial number of R microorganisms. Indeed, noticing that 
\begin{equation}
p_R^e(i)=\lim_{t\to\infty}P(0,t|i,0)=\lim_{t\to\infty}\phi_{i}(0,t)
\label{pRe}
\end{equation} 
will enable us to calculate $p_R^e(i)$~\cite{Alexander12,Parzen}. 

The probability $P(j,t|i,0)$ of having $j$ R mutants at time $t$, starting from $i$ R mutants at time $t=0$, satisfies the master equation
	\begin{equation}
	\frac{\partial P(j,t|i,0)}{\partial t}=f_R(t)\left(j-1\right)P(j-1,t|i,0)+g_R\left(j+1\right)P(j+1,t|i,0)-(f_R(t)+g_R)\,j\,P(j,t|i,0)\mbox{ }.
	\label{ME_tdbdp}
	\end{equation}
Here, we neglect mutants that appear after the addition of antimicrobial, and we deal with them in the calculation of $p_R^{a}$ and $p_R^{e'}$. The generating function defined in Eq.~\ref{PGF} satisfies the partial differential equation 
\begin{equation}
\frac{\partial \phi_{i}(z,t)}{\partial t}-(z-1)(f_R(t)z-g_R)\frac{\partial \phi_{i}(z,t)}{\partial z}=0\,.
\label{pdepgf}
\end{equation}
This first-order nonlinear partial differential equation can be solved using the method of characteristics. For this, we rewrite it as:
\begin{equation}
\vec{v}.\vec{\nabla}\phi_{i}=0 \mbox{ },
\end{equation}
where $\vec{v}=(1,\,\, -(z-1)(f_B(t)z-g_B))^t$ and $\vec{\nabla}\phi_{i}=(\partial \phi_{i}/\partial t,\,\, \partial \phi_{i}/\partial z)^t$. A characteristic curve $\vec{r}(s)$ satisfies $d\vec{r}/ds=\vec{v}(\vec{r}(s))$, which entails
\begin{equation}
\frac{d\phi_{i}}{ds}=\frac{d\vec{r}}{ds}.\vec{\nabla}\phi_{i}=\vec{v}.\vec{\nabla}\phi_{i}=0\,,
\label{phicst}
\end{equation}
implying that $\phi_i$ is constant along a characteristic curve.
Since $d \phi_{i}/ds=(\partial\phi_{i}/\partial t)(dt/ds)+(\partial\phi_{i}/\partial z)(dz/ds)$, we obtain the following system of ordinary differential equations along a characteristic curve:
\begin{equation}
\begin{cases}
\frac{dt}{ds}=1 \,,\\
\frac{dz}{ds}=-(z-1)(f_R(t)z-g_R)\,.
\end{cases}
\end{equation}
We choose to integrate it as
\begin{equation}
\begin{cases}
t=s\,, \\
\frac{dz}{dt}=-(z-1)(f_R(t)z-g_R)\,.
\end{cases} 
\end{equation}
The second ordinary differential equation can be solved by introducing $y=1/(z-1)$, which yields
\begin{equation}
\frac{e^{\rho(t)}}{z-1}-\int_{0}^tf_R(u)e^{\rho(u)}du=\frac{1}{z_0-1}\,,
\label{char}
\end{equation}
with
\begin{equation}
\rho(t)=\int_{0}^t\left(g_R-f_R(u)\right)du\,,
\label{rho}
\end{equation} 
where we have employed Eqs.~\ref{det1} and~\ref{det2}. Eq.~\ref{char} is the equation of the characteristic line going through the point $(0,\,z_0)$. Because $\phi_i$ is constant along this line (see Eq.~\ref{phicst}), we have $\phi_{i}(z,t)=\phi_{i}(z_0,0)=z_0^{i}$ along this line, where we have used Eq.~\ref{PGF}. Furthermore, for any $(z,t)$ we can find the appropriate $z_0$ using Eq.~\ref{char}. This yields the following expression for the generating function:
\begin{equation}
\phi_{i}(z,t)=\left[1+\left(\frac{e^{\rho(t)}}{z-1}-\int_0^tf_R(u)e^{\rho(u)}du\right)^{-1}\right]^{i}\,,
\label{phiexpl}
\end{equation}
where $\rho(t)$ is given by Eq.~\ref{rho} and $f_R(t)$ by Eq.~\ref{det1}.

We can now express the probability $p_R^e(i)$ from Eqs.~\ref{pRe} and~\ref{phiexpl}:
\begin{equation}
p_R^e(i)=\lim_{t\to\infty}\left[\frac{g_R\int_0^{t}e^{\rho(u)}du}{1+g_R\int_0^{t}e^{\rho(u)}du}\right]^{i}\,.
\label{pRe_f}
\end{equation}


\subsection{Predicting the extinction probability $p_0$}

\begin{figure}[htb]%
	\centering
	\includegraphics[width=0.885\textwidth]{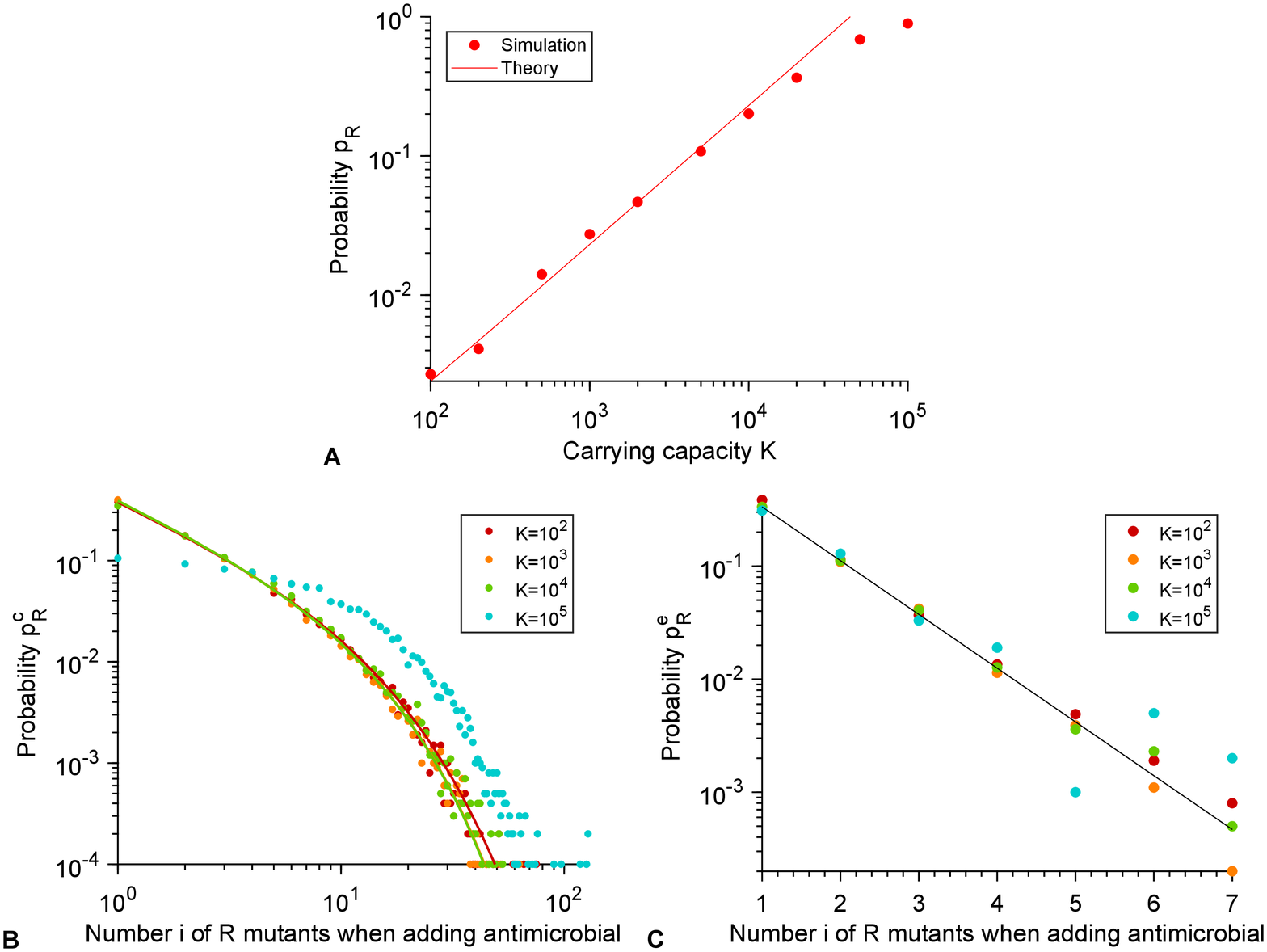}
	\vspace{0.2cm}
	\caption{{\bf Perfect biostatic antimicrobial: test of analytical predictions for each term involved in $p_0$ (Eq.~\ref{p0_biostatic}).}
		{\bf A:} Probability $p_R$ that at least one R mutant is present when antimicrobial is added, plotted versus carrying capacity $K$. Markers: simulation results, with probabilities estimated over $10^4$ realizations. Red solid line: analytical prediction, $p_R=t_R^{app}/\tau_R^d=N\mu_1 g_S \tau_R^d$ (see main text). {\bf B:} Probability $p_R^c$ that exactly $i$ R microorganisms are present when antimicrobial is added, provided that at least one R mutant is present, plotted versus the number $i$ of R mutants, for various carrying capacities $K$. Markers: simulation results, estimated over $10^4$ realizations. Solid lines: analytical prediction in Eq.~\ref{pRci}. Analytical prediction lines for $K=10^4$ and $K=10^5$ are confounded; note that the prediction holds in the weak mutation regime $K\mu_1\ll 1$, and thus fails for $K=10^5$ here. {\bf C:} Probability $p_R^e$ of rapid extinction of the R lineage, plotted versus the number $i$ of R mutants present when adding antimicrobial, for various different carrying capacities $K$. Markers: simulation results, with probabilities estimated over $10^4$ realizations. Black solid line: analytical prediction from Eq.~\ref{pRe_bios} (see main text). Parameter values: $f_S=1$ without antimicrobial, $f_S'=0$ with antimicrobial, $f_R=0.9$, $g_S=g_R=0.1$ and $\mu_1=10^{-5}$ ({\bf A}-{\bf B}) or $\mu_1=0$ ({\bf C}). }%
	\label{pR_pRc_pRe}%
\end{figure}

\begin{figure}[htb]%
	\centering
	\includegraphics[width=0.92\textwidth]{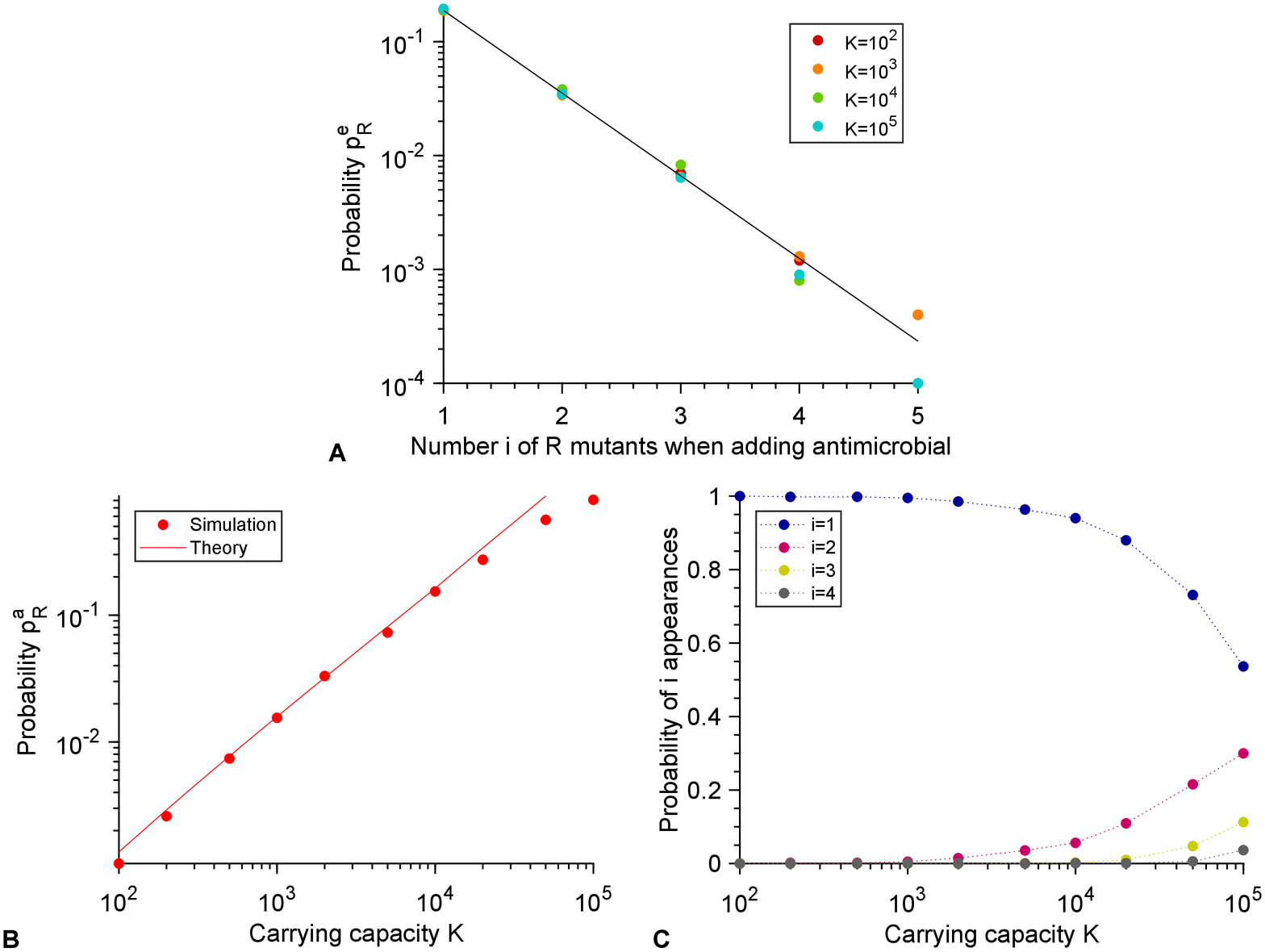}
	\vspace{0.2cm}
	\caption{{\bf Biocidal antimicrobial: test of analytical predictions for each term involved in $p_0$ (see Eq.~\ref{p0_biocidal}).} Note that $p_R$ and $p_R^c$ are the same as in Fig.~\ref{pR_pRc_pRe}A-B. {\bf A:} Probability $p_R^e$ of rapid extinction of the R lineage, plotted versus the number $i$ of R mutants present when adding antimicrobial, for various different carrying capacities $K$. Markers: simulation results, with probabilities estimated over $10^4$ realizations. Black solid line: analytical prediction from Eq.~\ref{pRe_f}. {\bf B:} Probability $p_R^a$ that resistance appears in the presence of antimicrobial, plotted versus the carrying capacity $K$. Red markers: simulation results, with probabilities estimated over $10^4$ realizations. Red solid line: analytical prediction, $p_R^a=N_{div}\mu_1$ with $N_{div}$ in Eq.~\ref{Ndiv}. {\bf C:} Probability that $i$ distinct lineages of R mutants appear in the presence of antimicrobial, provided that at least one appears, plotted versus the carrying capacity $K$. Markers: simulation results, with probabilities estimated over $10^3$ realizations. Parameter values: $f_S=1$, $f_R=0.9$, $g_S=0.1$ without antimicrobial, $g_S'=1.1$ with antimicrobial, $g_R=0.1$ and $\mu_1=0$ (panel {\bf A}) or $\mu_1=10^{-5}$ (panels {\bf B} and {\bf C}).}%
	\label{pRe_pRa_pi}%
\end{figure}

Here, we test the analytical predictions for each term involved in the extinction probability $p_0$ of the population above the MIC, both in the perfect biostatic case (see Eq.~\ref{p0_biostatic}) and in the biocidal case (see Eq.~\ref{p0_biocidal}), by comparing them to numerical simulation results. To estimate the probability $p_R$ that at least one R mutant is present when antimicrobial is added, and to study the number of R mutants that are then present (Fig.~\ref{pR_pRc_pRe}A-B), simulations are run starting from $j_0=10$ S microorganisms (and no R) as in the rest of our work. We let the population evolve until a specific time, in practice $t=500$, when population size is well-equilibrated around the deterministic stationary value $K(1-g_S/f_S)$ without antimicrobial, and we analyze population composition at this time. To estimate the probability $p_R^e$ of rapid extinction of the R lineage (Figs.~\ref{pR_pRc_pRe}C and~\ref{pRe_pRa_pi}A), we start from a population with $i$ R microorganisms and $K(1-g_S/f_S)-i$ sensitive microorganisms, and we let it evolve with antimicrobial until extinction of the S microorganisms. All these simulations are run with 2 types of microorganisms, S and R (no compensation). In Figs.~\ref{pR_pRc_pRe}C and~\ref{pRe_pRa_pi}A, we note that $p_R^e$ does not seem to depend on $K$. In fact, our analytical estimate for $p_R^e$ is fully independent of $K$ because it only involves the ratio $S(t)/K$ (see Eqs.~\ref{pRe_f},~\ref{rho} and~\ref{det1}), whose deterministic dynamics is independent of $K$ (see Eq.~\ref{edo} with $N(t)\leftarrow S(t)$).

\begin{figure}[htb]%
	\centering
	\includegraphics[width=0.92\textwidth]{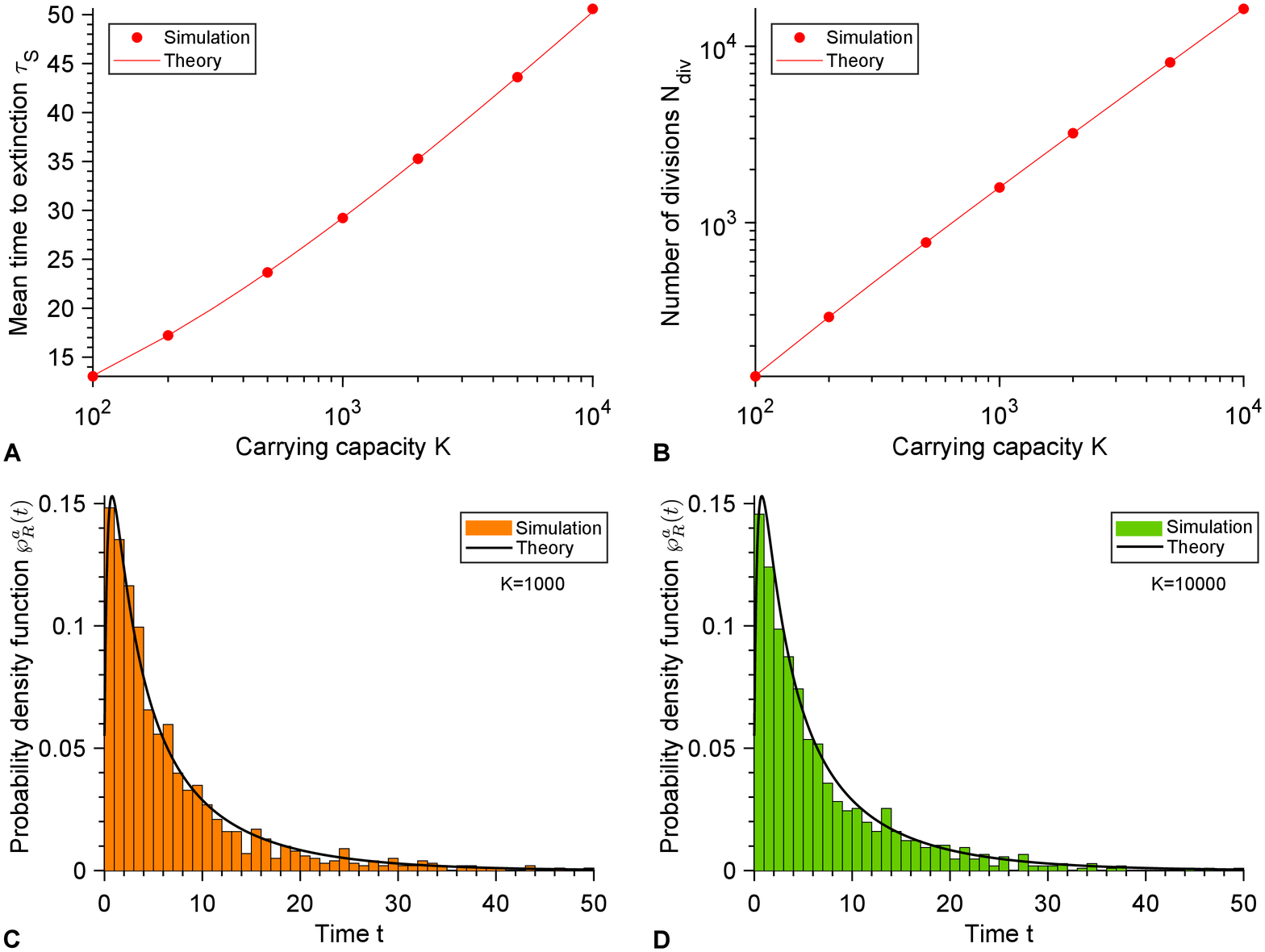}
	\vspace{0.2cm}
	\caption{{\bf Biocidal antimicrobial: test of analytical predictions for intermediate quantities involved in the calculation of $p_0$ (see Eq.~\ref{p0_biocidal}).} {\bf A:} Average time $\tau_S$ to extinction of a population of S microorganisms in the presence of antimicrobial, plotted versus the carrying capacity $K$. Markers: simulation results, with probabilities estimated over $10^4$ realizations. Red solid line: analytical prediction from Eq.~\ref{tj0}, with $j_0=K(1-g_S/f_S)$. {\bf B:} Number $N_{div}$ of individual division events that occur between the addition of antimicrobial and the extinction of the population of S microorganisms, plotted versus carrying capacity $K$. Red markers: simulation results, with probabilities estimated over $10^4$ realizations. Red solid line: analytical prediction from Eq.~\ref{Ndiv}. {\bf C and D:} Probability density function $\wp_R^a(t)$ of the time $t$ of appearance of an R mutant, under the assumption that exactly one R mutant appears between the addition of antimicrobial and the extinction of the population of S microorganisms, for $K=10^3$ ({\bf C}) and $K=10^4$ ({\bf D}). Histograms: simulation results, with $10^3$ realizations. Black solid lines: analytical prediction from Eq.~\ref{pdf_pRa}. Parameter values: $f_S=1$, $g_S=0.1$ without antimicrobial, $g_S'=1.1$ with antimicrobial, and in panels {\bf C} and {\bf D}, $f_R=0.9$, $g_R=0.1$ and $\mu_1=10^{-5}$.}%
	\label{tauS_Ndiv_pdf}%
\end{figure}

The probability $p_R^a$ that resistance appears in the presence of antimicrobial involves the number of divisions $N_{div}$ and the mean time to extinction $\tau_S$ of a population of S microorganisms in the presence of antimicrobial (see main text). To estimate these two intermediate quantities, simulations only involving S microorganisms in the presence of antimicrobial, starting from $K(1-g_S/f_S)$ sensitive microorganisms, are performed (Fig.~\ref{tauS_Ndiv_pdf}A-B). For $p_R^a$ itself (Fig.~\ref{pRe_pRa_pi}B), simulations with S and R microbes (no compensation), also starting from $K(1-g_S/f_S)$ sensitive microorganisms in the presence of antimicrobial, are performed. The time of appearance of R mutants (Fig.~\ref{tauS_Ndiv_pdf}C-D) and the number of different lineages that appear during the decay of this population (Fig.~\ref{pRe_pRa_pi}C) are also studied.

\clearpage

\subsection{A perfect biostatic antimicrobial yields a larger $p_0$ than a perfect biocidal antimicrobial}
\label{Perfectperfect}
For a perfect biostatic antimicrobial, the extinction probability $p_0$ upon the first addition of drug is given by Eq.~\ref{p0_biostatic}:
\begin{equation}
p_0=1-p_R\sum_{i=1}^{N-1}p_R^c(i)(1-p_R^e(i))\,,
 \label{p0_biostatic_bis}
\end{equation} 
while for a biocidal antimicrobial, the extinction probability $\tilde{p}_0$ upon the first addition of drug is given by Eq.~\ref{p0_biocidal}:
\begin{equation}
\tilde{p}_0=\left[1-p_R\sum_{i=1}^{N-1}p_R^c(i)(1-\tilde{p}_R^e(i))\right]\left[1-p_R^a(1-p_R^{e'})\right]<1-p_R\sum_{i=1}^{N-1}p_R^c(i)(1-\tilde{p}_R^e(i))\,.
 \label{p0_biocidal_bis}
\end{equation} 
In Eq.~\ref{p0_biocidal_bis} we have employed tilde symbols to denote the quantities that differ compared to Eq.~\ref{p0_biostatic_bis}. Recall that $p_R$ and $p_R^c(i)$ are the same in both cases. Indeed, these quantities characterize the state of the population when the antimicrobial is added, and thus do not depend on the type of treatment subsequently added.

The perfect biocidal antimicrobial corresponds to $g'_S\rightarrow\infty$. Let us prove that $\lim_{g'_S\rightarrow\infty}\tilde{p}_0<p_0$. From Eqs.~\ref{p0_biostatic_bis} and~\ref{p0_biocidal_bis} it is apparent that it suffices to prove that $\lim_{g'_S\rightarrow\infty}\tilde{p}_R^e(i)<p_R^e(i)$ for all $i$. The expression of both $p_R^e(i)$ and $\tilde{p}_R^e(i)$ is given in Eq.~\ref{pRe_f}, but it involves the decaying number $S(t)$ of S microorganisms once antimicrobial is added, which is different in these two cases, and is given respectively by Eq.~\ref{detg} with $f'_S=f_S$ in the biocidal case and by Eq.~\ref{det2} in the perfect biostatic case.

Taking the limit $g'_S\rightarrow\infty$ in Eq.~\ref{pRe_f} yields $\lim_{g'_S\rightarrow\infty}\tilde{p}_R^e(i)=(g_R/f_R)^i$, which corresponds to the extinction probability of a population that starts from $i$ R microorganisms, in the absence of any other microorganisms~\cite{Coates18}. But for a perfect biostatic antimicrobial,
\begin{equation}
\rho(t)=\int_0^t\left[g_R-f_R\left(1-\frac{S(u)}{K}\right)\right]du>\int_0^t\left[g_R-f_R\right]du=(g_R-f_R)t\,,
\end{equation}
which, using Eq.~\ref{pRe_f}, entails that $p_R^e(i)>(g_R/f_R)^i$, i.e. $\lim_{g'_S\rightarrow\infty}\tilde{p}_R^e(i)<p_R^e(i)$ for all $i$. Therefore, we have shown that $\lim_{g'_S\rightarrow\infty}\tilde{p}_0<p_0$: the extinction probability $p_0$ is larger for a perfect biostatic antimicrobial than for a perfect biocidal antimicrobial.

Importantly, our proof does not rely on the appearance of resistant microorganisms while antimicrobial is present, which cannot happen with a perfect biostatic antimicrobial, and whose probability tends to zero when $g'_S\rightarrow\infty$ with a biocidal antimicrobial. What makes the perfect biostatic antimicrobial more efficient than the perfect biocidal one is that S microorganisms survive for a longer time, thereby reducing the division rate of R microorganisms due to the logistic term, and favoring their extinction. Such a competition effect is realistic if S microorganisms still take up resources (e.g. nutrients) even while they are not dividing. 

\clearpage

\section{Fixation probability of a mutant in a population of constant size}
\label{Sec_PSR}

In the main text, in our discussion of sub-MIC concentrations of antimicrobials, we employed the fixation probability $p_{SR}$ of an R mutant in a population of S individuals with fixed size $N$:
\begin{equation}
p_{SR}=\frac{1-f_Sg_R/(f_Rg_S)}{1-[f_Sg_R/(f_Rg_S)]^{N}}\mbox{ }.
\label{PSR}
\end{equation}
Here, we briefly justify this formula. 

Consider a birth-death process in which, at each discrete time step, one individual is chosen with a probability proportional to its fitness to reproduce and another one is chosen with a probability proportional to its death rate to die. Note that at each time step, the total number of individuals in the population stays constant. This model is a variant of the Moran model with selection both on division and on death. Let $i$ be the number of R microorganisms and $N-i$ the number of S microorganisms. At a given time step, the probability $T_i^+$ that the number of R individuals increases from $i$ to $i+1$ satisfies:
\begin{equation}
T_i^+=\frac{f_Ri}{f_Ri+f_S(N-i)}\frac{g_S(N-i)}{g_Ri+g_S(N-i)}\mbox{ },
\end{equation}
and similarly, the probability $T_i^-$ that $i$ decreases by $1$ is given by:
\begin{equation}
T_i^-=\frac{f_S(N-i)}{f_Ri+f_S(N-i)}\frac{g_Ri}{g_Ri+g_S(N-i)}\mbox{ }.
\end{equation}
The probability $p_{SR}$ that the R genotype fixes in the population, starting from 1 R microorganism, then satisfies~\cite{Traulsen09}:
\begin{equation}
p_{SR}=\frac{1}{1+\sum_{k=1}^{N-1}\prod_{j=1}^k \gamma_j}\mbox{ },
\end{equation}
where
\begin{equation}
\gamma_i=\frac{T_i^-}{T_i^+}=\frac{f_Sg_R}{f_Rg_S}\mbox{ }.
\end{equation}
We thus obtain the result announced in Eq.~\ref{PSR}.

\clearpage

\section{Detailed simulation methods}
\label{SI_Simu}

In this work, the evolution of microbial populations are simulated using a Gillespie algorithm~\cite{Gillespie76,Gillespie77}. Let us denote by $j_S$, $j_R$ and $j_C$ the respective numbers of S, R and C individuals. The elementary events that can happen are division with or without mutation and death of an individual microbe of either type:
	\begin{itemize}
		\item $S \xrightarrow{k_S^{+}} 2S$: Reproduction without mutation of a sensitive microbe with rate $k_S^{+}=f_S^e(1-(j_S+j_R+j_C)/K)(1-\mu_1)$, with $f_S^e=f_S$ if no antimicrobial is present in the environment or $f_S^e=f'_S$ if antimicrobial is present in the environment. 
		\item $S \xrightarrow{k_{SR}} S+R$: Reproduction with mutation of a sensitive microbe with rate $k_{SR}=f_S^e(1-(j_S+j_R+j_C)/K)\mu_1$.
		\item $S \xrightarrow{k_S^{-}} \emptyset$: Death of a sensitive microbe with rate $k_S^{-}=g_S^e$, with $g_S^e=g_S$ if no antimicrobial is present in the environment or $g_S^e=g'_S$ if antimicrobial is present in the environment.
		\item $R \xrightarrow{k_R^{+}} 2R$: Reproduction without mutation of a resistant microbe with rate $k_R^{+}=f_R(1-(j_S+j_R+j_C)/K)(1-\mu_2)$.
		\item $R \xrightarrow{k_{RC}} R+C$: Reproduction with mutation of a resistant microbe with rate $k_{RC}=f_R(1-(j_S+j_R+j_C)/K)\mu_2$. 
		\item $R \xrightarrow{k_R^{-}} \emptyset$: Death of a resistant microbe with rate $k_R^{-}=g_R$.
		\item $C \xrightarrow{k_C^{+}} 2C$: Reproduction of a resistant-compensated microbe with rate $k_C^{+}=f_C(1-(j_S+j_R+j_C)/K)$. 
		\item $C \xrightarrow{k_C^{-}} \emptyset$: Death of a resistant-compensated microbe with rate $k_C^{-}=g_C$.
	\end{itemize}
	The total rate of events is given by $k_{tot}=(k_S^++k_{SR}+k_S^-)j_S+(k_R^++k_{RC}+k_R^-)j_R+(k_C^++k_C^-)j_C$.\\

Simulation steps are as follows:
\begin{enumerate}
	\item Initialization: The microbial population starts from $j_S=10$ sensitive microorganisms, $j_R=0$ resistant mutant and $j_C=0$ resistant-compensated mutant at time $t=0$ without antimicrobial. The next time when the environment changes is stored in the variable $t_{switch}$, which is initialized at $t_{switch}=T/2$, the first time when antimicrobial is added.
	\item The time increment $\Delta t$ is sampled randomly from an exponential distribution with mean $1/k_{tot}$, and the next event that may occur is chosen randomly, proportionally to its probability $k/k_{tot}$, where $k$ is its rate. For instance, division of a sensitive microorganism without mutation is chosen with probability $k_S^+ j_S/k_{tot}$.
	\item If $t+\Delta t < t_{switch}$, time is increased to $t+\Delta t$ and the event chosen at Step 2 is executed.
	\item If $t+\Delta t \geq t_{switch}$, the event chosen at Step 2 is not executed, because an environment change has to occur before. The environment change is performed: time is incremented to $t=t_{switch}$, and the fitness and death rate of the sensitive microbes are switched from $f_S$ to $f'_S$ and from $g_S$ to $g'_S$ or vice-versa. In addition, $t_{switch}$ is incremented to $t_{switch}+T/2$, and thus stores the next time when the environment changes. 
	\item We go back to Step 2 and iterate until the total number of microbes is zero ($j_S+j_R+j_C=0$) or there are only resistant-compensated mutants ($j_S=0$, $j_R=0$ and $j_C \neq 0$).
\end{enumerate}

Note that Step 4 introduces an artificial discretization of time, because environment changes occur at fixed times and not with a fixed rate. However, because the total event rate is large unless the population size is very small, the ``jump" in time induced by Step 4 is usually extremely small, and the discarded events constitute a tiny minority of events. The resulting error is thus expected to be negligible. The very good agreement between our simulation results and our analytical predictions, in particular for short periods, corroborates this point.



\end{document}